\documentclass[fleqn,usenatbib]{mnras}

\usepackage{mathptmx}

\usepackage[T1]{fontenc}

\usepackage{graphicx}	
\usepackage{amsmath}	
\usepackage{amssymb}	
\usepackage[usenames,dvipsnames]{xcolor} 
\usepackage[normalem]{ulem} 
\usepackage{hyperref,siunitx}

\newcommand{\eagle}{\mbox{\sc{Eagle}}}
\newcommand{\euclid}{\mbox{\it Euclid}}
\newcommand{\hst}{\mbox{\it HST}}
\newcommand{\jwst}{\mbox{\it JWST}}
\newcommand{\flares}{\mbox{\sc Flares}}
\newcommand{\flare}{\mbox{\sc Flare}}
\newcommand{\rst}{\mbox{\it Roman Space Telescope}}
\newcommand{\alma}{\mbox{\textit{ALMA}}}

\newcommand{\skirt}{\mbox{\sc skirt}}

\newcommand{\eg}[0]{$\textnormal{e.g. }$}

\newcommand{\etc}[0]{$\textnormal{etc}$}
\newcommand{\mumetre}{\SI{}{\micro\meter}}

\definecolor{deepchestnut}{rgb}{0.73, 0.31, 0.28}

\title[FLARES III]{First Light And Reionisation Epoch Simulations \\(FLARES) III: The properties of massive dusty galaxies at cosmic dawn}

\author[A.P. Vijayan et al.]{Aswin P. Vijayan$^{1,2,3}$\thanks{E-mail: apavi@space.dtu.dk},
Stephen M. Wilkins$^{1}$, 
Christopher C. Lovell$^4$,
Peter A. Thomas$^1$,
\newauthor
Peter Camps$^5$, 
Maarten Baes$^5$, 
James Trayford$^6$,
Jussi Kuusisto$^{1}$,  
William J. Roper$^{1}$ 
\\
$^{1}$Astronomy Centre, University of Sussex, Falmer, Brighton BN1 9QH, UK\\
$^{2}$Cosmic Dawn Center (DAWN)\\
$^{3}$DTU-Space, Technical University of Denmark, Elektrovej 327, DK-2800 Kgs. Lyngby, Denmark\\
$^{4}$Centre for Astrophysics Research, School of Physics, Astronomy $\&$ Mathematics, University of Hertfordshire, Hatfield AL10 9AB, UK\\
$^{5}$Sterrenkundig Observatorium, Universiteit Gent, Krijgslaan 281 S9, 9000 Gent, Belgium\\
$^{6}$Institute of Cosmology \& Gravitation, University of Portsmouth, Dennis Sciama Building, Burnaby Road, Portsmouth PO1 3FX, UK
}

\date{Accepted XXX. Received YYY; in original form ZZZ}

\pubyear{2022}

\begin{document}
\label{firstpage}
\pagerange{\pageref{firstpage}--\pageref{lastpage}}
\maketitle

\begin{abstract}
Using the First Light And Reionisation Epoch Simulations (\textsc{Flares}), we explore the dust-driven properties of massive high-redshift galaxies at $z\in[5,10]$. By post-processing the galaxy sample using the radiative transfer code \textsc{skirt} we obtain the full spectral energy distribution. We explore the resultant luminosity functions, IRX-$\beta$ relations as well as the luminosity-weighted dust temperatures in the Epoch of Reionisation (EoR). We find that most of our results are in agreement with the current set of observations, but underpredict the number densities of bright IR galaxies, which are extremely biased towards the most overdense regions. We see that the \textsc{Flares} IRX-$\beta$ relation (for $5\le z\le8$) pre-dominantly follows the local starburst relation. The IRX shows an increase with stellar mass, plateauing at the high-mass end ($\sim10^{10}$ M$_{\odot}$) and shows no evolution in the median normalisation with redshift. We also look at the dependence of the peak dust temperature ($T_{\mathrm{peak}}$) on various galaxy properties including the stellar mass, IR luminosity and sSFR, finding the correlation to be strongest with sSFR. The luminosity-weighted dust temperatures increase towards higher redshifts, with the slope of the $T_{\mathrm{peak}}$ - redshift relation showing a higher slope than the lower redshift relations obtained from previous observational and theoretical works. The results from \textsc{Flares}, which are able to provide a better statistical sample of high-redshift galaxies compared to other simulations, provides a distinct vantage point for the high-redshift Universe. 
\end{abstract}




\section{Introduction}\label{sec:intro}
The \textit{Hubble Space Telescope} (\hst) has been instrumental in the last decade observing the rest-frame UV of high-redshift galaxies \cite[\eg][]{Beckwith_2006,Bouwens2006,Bunker2010,Grogin2011,Salmon2020}, finding more than 1000 galaxies at $z>5$. These efforts from \hst\ have been complemented by wide-area ground based near-IR surveys \cite[\eg \textit{UltraVISTA}, ][]{Bowler2014,Stefanon2019} providing samples of rare bright galaxies. \textit{Spitzer Space Telescope} observations \citep[\eg][]{Ashby_2013,Roberts_Borsani_2016}, probing the rest-frame optical at $z>5$ has provided further contraints on these high-redshift systems. However, the UV/optical alone cannot unravel the nature as well as dynamical properties of these high-redshift systems, such as reliable estimates of the total star formation rates, since it is not an unbiased tracer due to the presence of dust. 

Dust plays a major role in the observation of galaxies, with nearly half of all photons in the Universe reprocessed by dust grains during their lifetime \cite[\eg][]{Savage1979,Dwek1998,Bernstein2002,Driver2016}. Even though the average dust content of galaxies in the EoR is expected to be lower compared to the low-redshift Universe \cite[$z<4$, \eg][]{Li2019,Magnelli2020,Pozzi2021}, it still has a significant impact on shaping observations by attenuating the emitted source radiation \cite[see review by][]{Salim2020}, particularly on the most massive galaxies. Recent observations have also found dust-obscured galaxies \cite[as early as $z\sim7$, \eg][]{Fudamoto2021}, which were undetected in the rest-frame UV. Thus it is important that we probe the nature of dust and its affects to better understand the various observationally derived quantities of galaxies in the EoR. The stellar emission in a galaxy, which is predominantly in the UV-to-NIR, gets re-processed by the intervening dust into the IR regime. Over the years, the observations in this regime using far infrared (FIR), millimetre (mm) and sub-millimetre (sub-mm) observatories have been instrumental in mapping the dust content of galaxies. This has been done with the help of instruments like \alma\ \cite[\eg][]{Knudsen2017,Hashimoto2018,Smit2018,Bouwens2020}, \textit{Herschel} \cite[\eg][]{Gruppioni2013,Wang2019}, \etc{}. In many cases there have been detections from deep \alma\ and \textit{PdBI} observations of galaxies at extremely high redshifts ($z>6$) with large reservoirs of dust  \citep[$>10^8$ M$_{\odot}$;][]{Mortlock2011,Venemans2012,daCunha2015}.

The early identification of these dusty star-forming galaxies were from single-dish sub-mm surveys finding massive populations at $z>1$ \cite[see][]{Casey2014}. Even though they are rare, these galaxies contribute significantly to the cosmic star formation density during cosmic noon ($z\sim2-3$). The picture at higher redshift ($z>4$) is still unclear. \alma\ and \textit{Herschel} have been instrumental in filling this space at high-redshift. Recent survey programmes like ALPINE \cite[]{LeFevre2020,Bethermin2020,Faisst2020a} and ASPECs \cite[]{Walter2016,Decarli2019,Gonzalez-Lopez2019} are helping us to understand the dusty nature of high-redshift galaxies \cite[also see][for more high-z surveys]{Hodge2020} by building a large statistical sample. High-redshift studies like \cite{Gruppioni2013,Wang2019,Gruppioni2020} have constructed IR luminosity functions. The jury is still out on the normalisation of the IR LF due to difficulties in de-blending of IR data and smaller volumes probed in some surveys. Other observational studies like \cite{Schreiber2018} and \cite{Bouwens2020} have explored the evolution of the luminosity-weighted dust temperatures, and have found an increase in the value with increasing redshift (up to $z\le5$). Another important observational space that has been studied widely is the relationship between the Infrared Excess (IRX) - UV continuum slope ($\beta$). Empirical relationships \cite[\eg][]{Pettini1998,Meurer1999,Takeuchi2012,Reddy2015} built in this space using observations of low-redshift (mainly $z\lesssim2$) galaxies have been used to correct for dust attenuation in galaxies. There has been a variety of observational studies exploring this space at these high redshifts \cite[\eg][]{Koprowski2018,Fudamoto2020,Bouwens2020,Schouws2021}. They have found varying results that favours empirical relation using Calzetti as well as SMC extinction curve. The obtained relation is also strongly influenced by the adopted SED dust temperature and the functional form used to obtain the IR luminosity. With upcoming surveys on facilities like the \textit{James Webb Space Telescope} (\jwst), \euclid{}, \rst{}, and the \textit{Atacama Large Aperture Submillimeter Telescope} (\textit{AtLAST}, \citealt{Klaassen2019}) are expected to substantially contribute to these efforts to build a comprehensive picture of galaxy formation and evolution in the high-redshift Universe. 

In addition to these observational efforts, it is crucial to study the nature of these dusty high-redshift systems using theoretical models of galaxy formation and evolution.
Several studies have used techniques built with semi-analytical or analytical \cite[\eg][]{Lacey2016,Popping2017,Lagache2018,Lagos2019,Sommovigo2020} and hydrodynamical \cite[\eg][]{Olsen2017,Narayanan2018,Ma2019,McAlpine2019,Liang2019,Baes2020,Trcka2020,Lovell2021b,Shen2022} models to explore and understand the trends and variations in observed properties of galaxies like the dust temperatures, fine-structure transitions, infrared luminosity functions, IRX-$\beta$ relations, submillimeter number counts, \etc. Many of them have been successful in reproducing various observational results, and has also been instrumental in understanding the underlying scaling relations.  

In comparison to semi-analytical models, hydrodynamical simulations model in greater detail the evolution of dark matter, gas, stars, and black holes, allowing for a more detailed exploration of galaxy structure and observed properties. 
A drawback of some of the current state-of-the-art cosmological simulation periodic boxes [\eg \textsc{MassiveBlack} \citep{Di_Matteo_2012}, \eagle\, \citep{schaye2015_eagle,crain2015_eagle}, \textsc{BlueTides} \citep{Feng2016}, \textsc{Illustris-Tng} \citep{Naiman2018,Nelson2018,Marinacci2018,Springel2018,Pillepich2018}, \textsc{Simba} \citep{Dave2019}, \textsc{Cosmic Dawn ii} \citep{Ocvirk2020}, \etc] is that they have fewer massive galaxies in the EoR, which are thought to be biased towards the most overdense regions \cite[see][]{Chiang2013,Lovell2018,Ito2020}. 
This is mainly due to the unfeasible amount of computational time required to run much larger periodic volumes with the needed resolution to resolve the relevant scales at these high redshifts. 
Hence, they lack the statistical power to investigate the bright galaxies that will be discovered and investigated with the current or future generation of telescopes. 

To overcome this dearth of a representative statistical sample of massive galaxies when studying the EoR, we have run a suite of zoom simulations, the First Light And Reionisation Epoch Simulations, \flares{}; introduced in \cite{Lovell2021a,Vijayan2021} to re-simulate a wide range of overdensities. The simulations were run using the well studied  \eagle\ \cite[]{schaye2015_eagle,crain2015_eagle} model, which has been shown to be in good agreement with a large number observables at low and high redshift regime not used in the calibration \cite[\eg][]{furlong_evolution_2015,Lagos2015,Trayford2017}. 
In this paper, using the radiative transfer code \skirt\ \cite[]{skirt2015} we post-process the simulation to study the spectral energy distributions (SEDs) of massive galaxies in the EoR. 
We aim to understand how well the \eagle\ physics model is able to reproduce the high-redshift Universe, mostly in comparison to observations in the infrared part of the spectrum like the luminosity functions, IRX-$\beta$ and luminosity-weighted dust temperatures. 
This work complements other theoretical studies in the high-redshift Universe and provide insights into how the intrinsic galaxy properties are connected to their observed and derived properties. 

This paper is structured as follows, in section \S\ref{sec:methods} we introduce the suite of simulations, our galaxy sample and the method for SED generation. In section \S\ref{sec:res} we show our results, including the UV and IR luminosity in \S\ref{sec:res::LF}, the IRX-$\beta$ relation in \S\ref{sec:res::IRXbeta}, and the variation and evolution of dust temperatures in \S\ref{sec:res::Td}. 
We finally summarise our findings and present our conclusions in section \S\ref{sec:conc}. 
Throughout we assume a Planck year 1 cosmology \citep[$\Omega_{m}$ = 0.307, $\Omega_{\Lambda}$ = 0.693, h = 0.6777;][]{planck_collaboration_2014}. 
\section{Methods}\label{sec:methods}
\subsection{The \flare\, simulations}\label{sec:methods::sims}
We use the First Light And Reionisation Epoch Simulations \cite[\flares{},][]{Lovell2021a,Vijayan2021}, a suite of 40 zoom simulations chosen from a 3.2 cGpc a side dark matter only box resimulated with full hydrodynamics using the \eagle\, \cite[][]{schaye2015_eagle,crain2015_eagle} simulation physics, down to $z=4.7$. The simulations were run with a heavily modified version of \textsc{P-Gadget-3}, which was last described in \cite{Springel2005a}, an N-Body Tree-PM smoothed particle hydrodynamics (SPH) code. 
The model uses the hydrodynamic solver collectively known as \textsc{Anarchy} \citep[described in][]{schaye2015_eagle,Schaller2015}, that adopts the pressure-entropy formulation described by \cite{Hopkins2013}, an artificial viscosity switch \citep{cullen_inviscid_2010}, and an artificial conduction switch \citep[\eg][]{price_modelling_2008}. The model includes radiative cooling and photo-heating \citep{Wiersma2009a}, star formation \citep{Schaye2008}, stellar evolution and mass loss \citep{Wiersma2009b}, black hole growth \citep{Springel2005b} and feedback from star formation \citep{DallaVecchia2012} and
AGN \citep{Springel2005b,B_and_S2009,Rosas-Guevara2015}. 
We use the AGNdT9 configuration of the \eagle\, simulation physics, which produces similar mass functions to the Reference model but better reproduces the hot gas properties of groups and clusters \citep{barnes_cluster-eagle_2017}. This configuration gives less frequent, more energetic AGN outbursts. 

\flares\ has an identical resolution to the \eagle\ reference run, with a dark matter and an initial gas particle mass of \mbox{m$_{\mathrm{dm}}=9.7\times10^6$ M$_{\odot}$} and m$_{\mathrm{g}}=1.8\times10^6$ M$_{\odot}$ respectively, and has a gravitational softening length of $2.66$ ckpc at $z\ge2.8$. The regions are selected at $z=4.7$ and have a radius of 14 cMpc/h, spanning a wide range of overdensities, from
$\delta=-0.479\to0.970$ \cite[see Table A1 in][]{Lovell2021a}. We have deliberately selected a large number of extreme overdense regions (16) to obtain a large sample of massive galaxies. To obtain a representative sample of the Universe, the galaxies in various overdensities can be combined using an appropriate weighting scheme. For a more detailed description of the simulation and weighting method we refer the readers to \cite{Lovell2021a}. Since the galaxies analysed in this study are sampled from the different regions, the mean or median results quoted are weighted values based on that scheme.

\subsection{Galaxy Identification and Selection}\label{sec:methods::galid}
\begin{figure}
	\centering
	\includegraphics[width=\columnwidth]{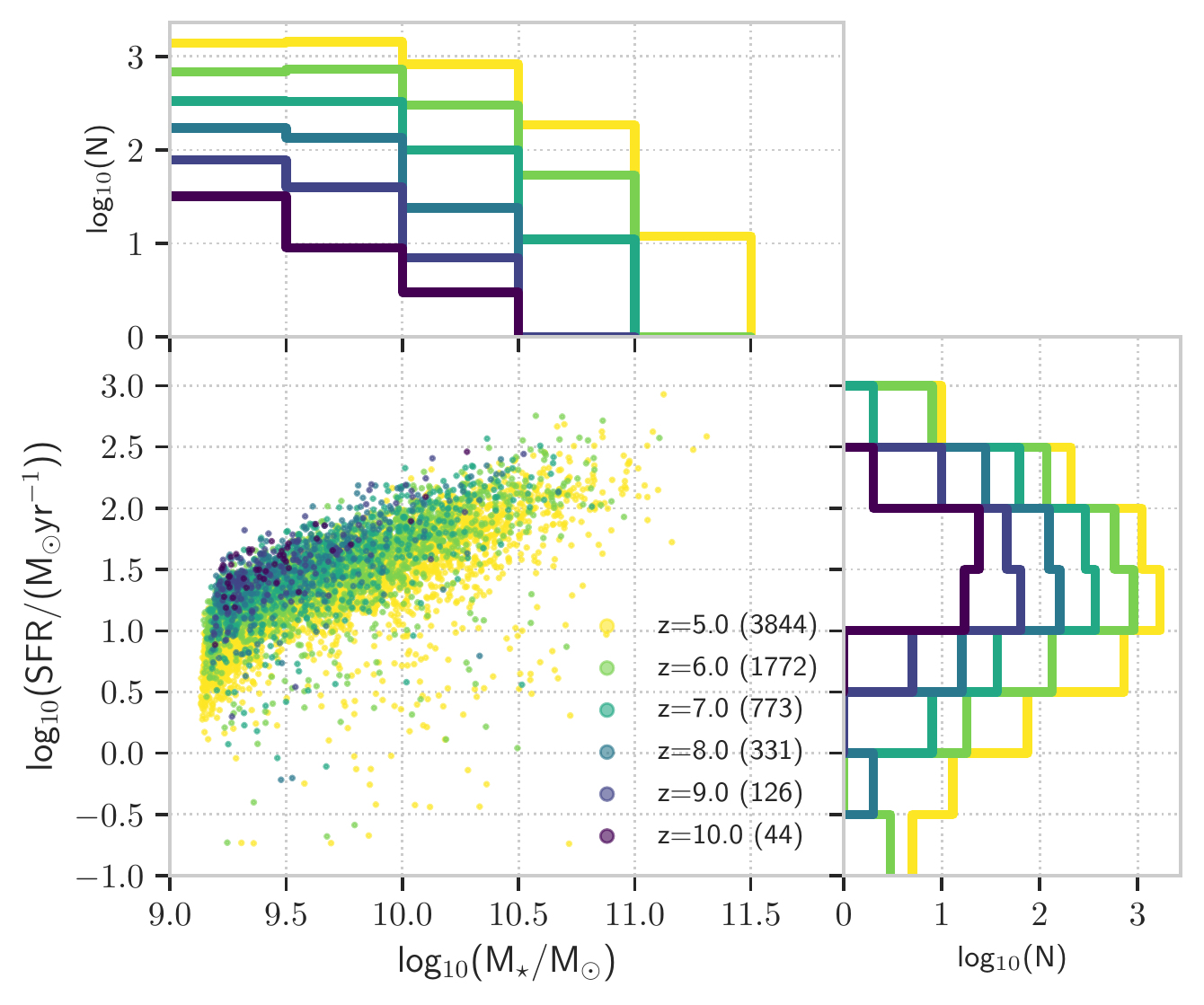}
	\caption{Shows the relationship between the galaxy stellar mass and the star formation rate (SFR) averaged over the star particles that were formed in the	last 10 Myr for $z\in [5, 10]$. Also shown is the histogram of the distribution of stellar mass and SFR in different bins for these redshifts. The total number of galaxies at these redshifts are indicated within brackets alongside the legend.\label{fig:Mstarsfr}}
\end{figure}

Galaxies in \flares, similar to the standard \eagle{}, are identified with the \textsc{Subfind} algorithm \cite[][]{Springel2001,Dolag2009}, which runs on bound groups found from via the Friends-Of-Friends algorithm \cite[\textsc{FoF},][]{Davis1985}. The galaxy stellar masses are defined using star particles within a 30 pkpc aperture centred on the most bound particle of the self-bound substructures. For the purpose of this study we concentrate only on the most well resolved galaxy systems that have more than 1000 star particles. This coincides with galaxies more massive than $\sim$10$^9$ M$_{\odot}$ in stellar mass (see Fig.~\ref{fig:Mstarsfr}). This selection also overlaps very well with the observationally inferred mass ranges \cite[for \eg the ALPINE survey;][]{LeFevre2020,Bethermin2020,Faisst2020a} of the galaxies detected in the infrared with ALMA and other instruments.

In Fig~\ref{fig:Mstarsfr} we plot the stellar mass of the selected galaxies against their star formation rate (SFR, quoted values are averaged for stars formed in the last 10 Myr) for $z\in[5,10]$. We also show histograms of the galaxy stellar masses and SFR distributions. 
Our selection samples $\sim 7000$ galaxies in this redshift and mass regime. 
At $z=10$, our sample of galaxies is only 44, and thus any inferences drawn can be subject to large scatter. 
The SFR seen in our selection has a maximum value just below $10^3$ M$_{\odot}$/yr. 

\subsection{Spectral Energy Distribution modelling}\label{sec:methods::sed}
\begin{figure}
	\centering
	\includegraphics[width=\columnwidth]{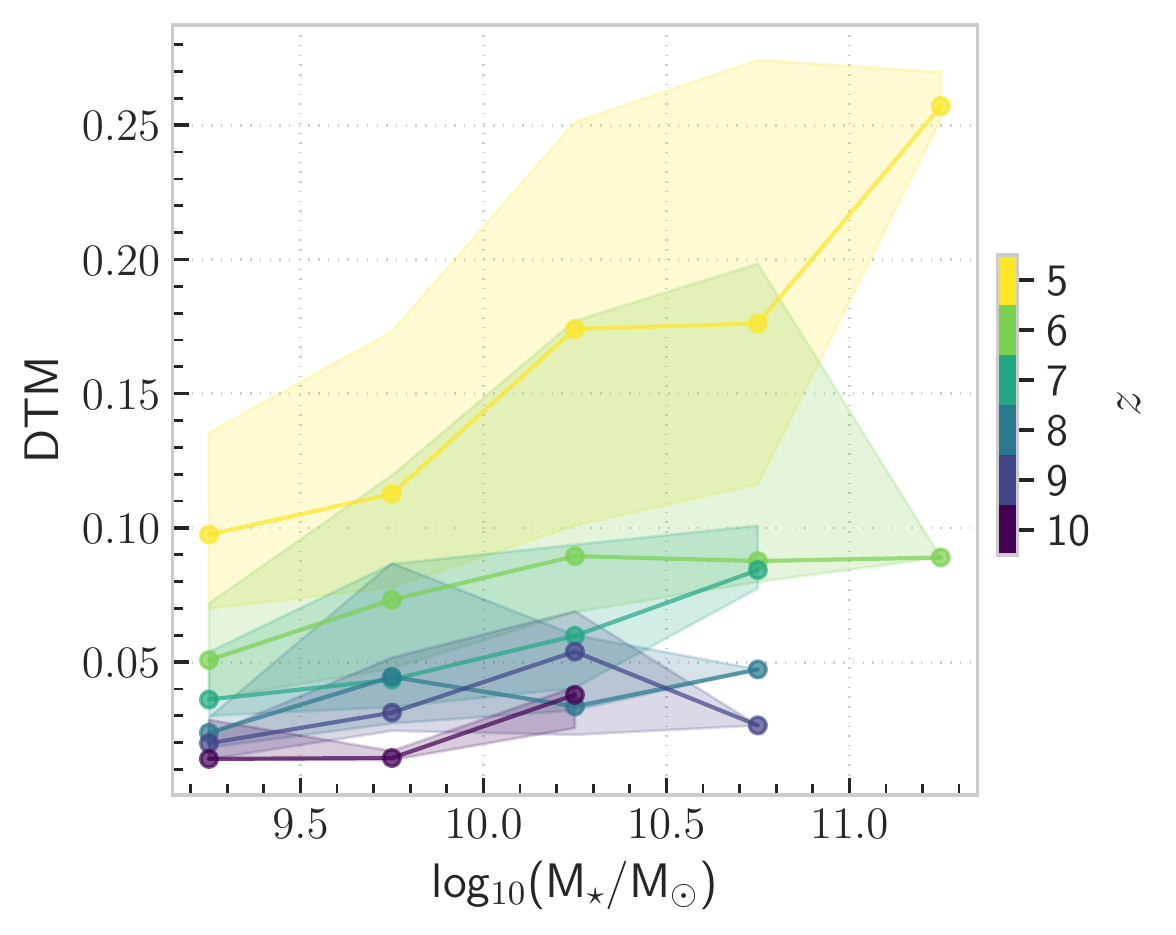}
	\caption{Shows the evolution of the dust-to-metal (DTM) ratio with the galaxy stellar mass across $z\in[5,10]$. The solid line shows the weighted median while the shaded region indicate the 16-84 percentile spread in the value. The DTM decreases with increasing redshift. \label{fig:DTM}}
\end{figure}

There are various methods to obtain the SED of a galaxy in simulations. A comprehensive method to get them, so that the properties of the dusty medium is captured, is to perform radiative transfer. 
There are numerous codes [\eg\ \textsc{Sunrise} \citep{Jonsson2006}, \textsc{Radmc-3d} \citep{RADMC}, \skirt\ \citep{skirt2015}, \textsc{Powderday} \citep{Narayanan2021}, \etc] available, most relying on sophisticated Monte Carlo methods. 
For this study we use the publicly available code \skirt{}, version 9 \cite[]{skirt9}.

\flares\, does not inherently model dust formation and destruction, and thus cannot reliably estimate the amount, nature and distribution of dust in the different galaxies. For the purpose of obtaining the amount and distribution of dust we assume a constant dust-to-metal ratio (DTM=M$_{\textrm{dust}}$/(M$_{\textrm{metal}}$ + M$_{\textrm{dust}}$)) per galaxy, in SPH gas particles below temperatures of 10$^6$ K or in star-forming gas particles. 
This temperature is higher than what was adopted in previous \eagle-\skirt\ work \cite[\eg][]{Camps2016,Trayford2017}, and ensures that dust is only destroyed in the very hot gas phase in the galaxies due to thermal sputtering \cite[see][for more details]{Draine1979,Tielens1994}. 
Changing the threshold to lower temperatures has negligible impact on the results presented in this work. 
The DTM ratio is calculated using the DTM fitting function in \citet[][Equation 15 in that work]{Vijayan2019}, obtained from the dust model implemented in the L-Galaxies semi-analytical model. In that work, the DTM ratio is parameterised as a function of the mass-weighted stellar age and the gas-phase metallicity. Fig.~\ref{fig:DTM} shows the evolution and spread in the DTM ratio used in this work as a function of the galaxy stellar mass for $z\in[5,10]$. It can be seen that there is an increase from a value of $\sim0.01$ at $z=10$ to $\sim0.2$ by $z=5$. 
The spread in the value also increases with decreasing redshift. More details on the evolution of the DTM ratio with redshift, and its dependence on various other galaxy properties, can be found in \cite{Vijayan2019}. The use of a varying DTM ratio dependent on galaxy properties, as opposed to a constant value of 0.3, is another difference from previous \eagle-\skirt\, work. 
The evolution of the median DTM ratio with redshift seen here is similar to the one observed in \cite{Vogelsberger2020} for \textsc{Illustris-Tng} galaxies.

In this work we use the Small Magellanic Cloud (SMC) dust grain type and size distribution \cite[]{Weingartner_Draine01} built in to \skirt. 
Due to its low-metallicity, the SMC is considered to be a good analogue to high-redshift galaxies. We use 8 grain size bins for silicate and graphite dust types to compute the thermal emission. The dust grid for this setup is constructed using the built-in octree grid in \skirt, using the previously defined dust particle distribution obtained from SPH gas particles. The octree is refined between a minimum refinement level of 6 and maximum of 16, with the cell splitting criterion set to a dust fraction value of 2$\times$10$^{-6}$ times the total dust mass in the domain, as well as a maximum V-band optical depth of 10. We use 10$^6$ photon packets per each radiation field wavelength grid, giving good convergence in observed properties. 
Our radiation field wavelength grid, as well as the dust emission grid, is spanned by a logarithmic grid between $0.08-1500$ \mumetre, with 200 points. We include dust self-absorption and re-emission in the set-up, with this procedure iterated such that the change in the absorbed dust luminosity is less than $3$ per cent. We place our detector to record the SED at a distance of 1 pMpc enclosing a 60 pkpc a side square region. 
In this work we record multiple orientation sightlines, but the fiducial orientation is along the z-axis.

Similar to previous \eagle-\skirt\ work, we apply differing amount of dust attenuation to old and young stellar populations. Young stars, with stellar ages less than 10$^7$ yr, are still embedded in their birth clouds, and as such experience higher dust attenuation \cite[\eg][]{Charlot_and_Fall}. 
We perform the same resampling technique that was employed in \cite{Camps2016,Trayford2017} to designate young and old stellar population from  star particles and star-forming gas particles in the simulation. 
A difference from those works is that we do not subtract the contribution of dust from young stars which were part of the star-forming gas particles when we perform the resampling. 
This is because these particles already have gas/dust intrinsic to them \cite[see section 2.4.4 in][about  introducing `ghost' gas particles]{Camps2016} unlike the resampled star particles which were converted to young stars. The
emission from old stellar populations is modelled using the BPASS \cite[]{BPASS2.2.1} SPS library and the young stars with MAPPINGS III \cite[]{Groves2008} templates. The former uses the Chabrier \cite[]{chabrier_galactic_2003} IMF while the latter uses Kroupa IMF \cite[]{Kroupa2002}. We do not expect this difference to have a big effect since they are very similar. The BPASS model is characterised by the age and metallicity of the stellar particle while the MAPPINGS III template uses the SFR, metallicity, the pressure of the ambient ISM, the compactness of the H\textsc{ii} region (log$_{10}$(C)), and the covering fraction of the associated photo-dissociation region ($f_{\mathrm{PDR}}$) of the star-forming particles. We use the same prescription for deriving the SFR, pressure and log$_{10}$(C) of the star particle as in \cite{Camps2016}. However, we set the PDR covering fraction, $f_{\mathrm{PDR}}$ to 0.2, higher than 0.1 which was used in \cite{Camps2016}. 
Our adopted value is same as the fiducial value used in \cite{Groves2008,Jonsson2010}. A higher $f_{\mathrm{PDR}}$ results in more of the stellar emission to be absorbed by the dust present within the birth-clouds, implying that more of the light is re-processed to the IR. Thus a higher value of $f_{\mathrm{PDR}}$ implies a higher value of the IR luminosity. However, it should be noted that some features of the galaxy SED (\eg change in the position where the IR flux density peaks) also depends on the value of log$_{10}$(C) \cite[also see \S5.4 in][]{Liang2019}. For more details about the \skirt\ set-up we have used, we refer the interested reader to \cite{Camps2016} and \cite{Trayford2017}.

We use the local thermal equilibrium set-up in \skirt\ which means that the dust grains are in local equilibrium with the radiation field. This condition (as opposed to being in non-thermal equilibrium) will pre-dominantly affect the fluxes in the rest-frame mid-IR, but have very negligible effect on our predictions in this work (for more details see Appendix~\ref{sec:conv}, where we have run \skirt\ with the non-thermal equilibrium setup). We also include dust heating from CMB radiation, which at high-redshifts \cite[since, T$_{\mathrm{CMB}}(z)=$T$_{\mathrm{CMB}}(z=0)\times(1+z)$; also see][]{daCunha2015} can be non-negligible. We do not include the effect of AGN on the SEDs (\skirt\ has the capability to model AGN emission, see \citealt{Stalevski2012,Stalevski2016}); we will show in Appendix~\ref{sec:AGN} how the predictions are affected when adding the AGN bolometric luminosity to the infrared emission (the effect is negligible and only seen at the bright IR luminosity end). 
In a future work (Kuusisto et al. in prep) we will explore in more detail the effect of AGN on the UV emission from \flares\ galaxies. 

We had previously modelled the UV to near-IR SED of the \flares\ galaxies in \cite{Vijayan2021} using a line-of-sight (LOS) dust extinction model. 
That work calibrated the dust attenuation based on matching to the UV luminosity function and the UV luminosity-UV-continuum slope relation at $z=5$, as well as the [O\textsc{iii}]$\lambda4959,5007+$H$\beta$ equivalent width relation at $z=8$. Here we do not perform any calibration, and only adopt the dust-to-metal ratio from the L-Galaxies SAM which was successful in reproducing many of the seen observational trends. This will enable us to better understand many of the successes and shortcomings of the \eagle\ model when applied at high-redshift. 
We compare the UV luminosity of the galaxies from this model to the LOS model in Appendix~\ref{sec:los_comp}.
\section{Results}\label{sec:res}
In this section we will look at what we can learn about the dust properties of massive high-redshift galaxies from the \flare\ simulations, focussing on $z\in[5,10]$. In \S\ref{sec:res::LF} we will look at the infrared (IR) luminosity function, while exploring the IRX-$\beta$ space in \S\ref{sec:res::IRXbeta}. In \S\ref{sec:res::Td} we will look at the dust temperatures of these galaxies, exploring both the SED--inferred as well as the peak dust temperatures. 
All these observables are also compared to the current observations. 
It should be noted that we do not model any observational effects (such as modelling the PSF or associated noise) that are inherent to the observed datasets that we compare to; this could impact derived properties and the associated systematic errors.

\subsection{Luminosity functions}\label{sec:res::LF}
\begin{figure*}
	\centering
	\includegraphics[width=\textwidth]{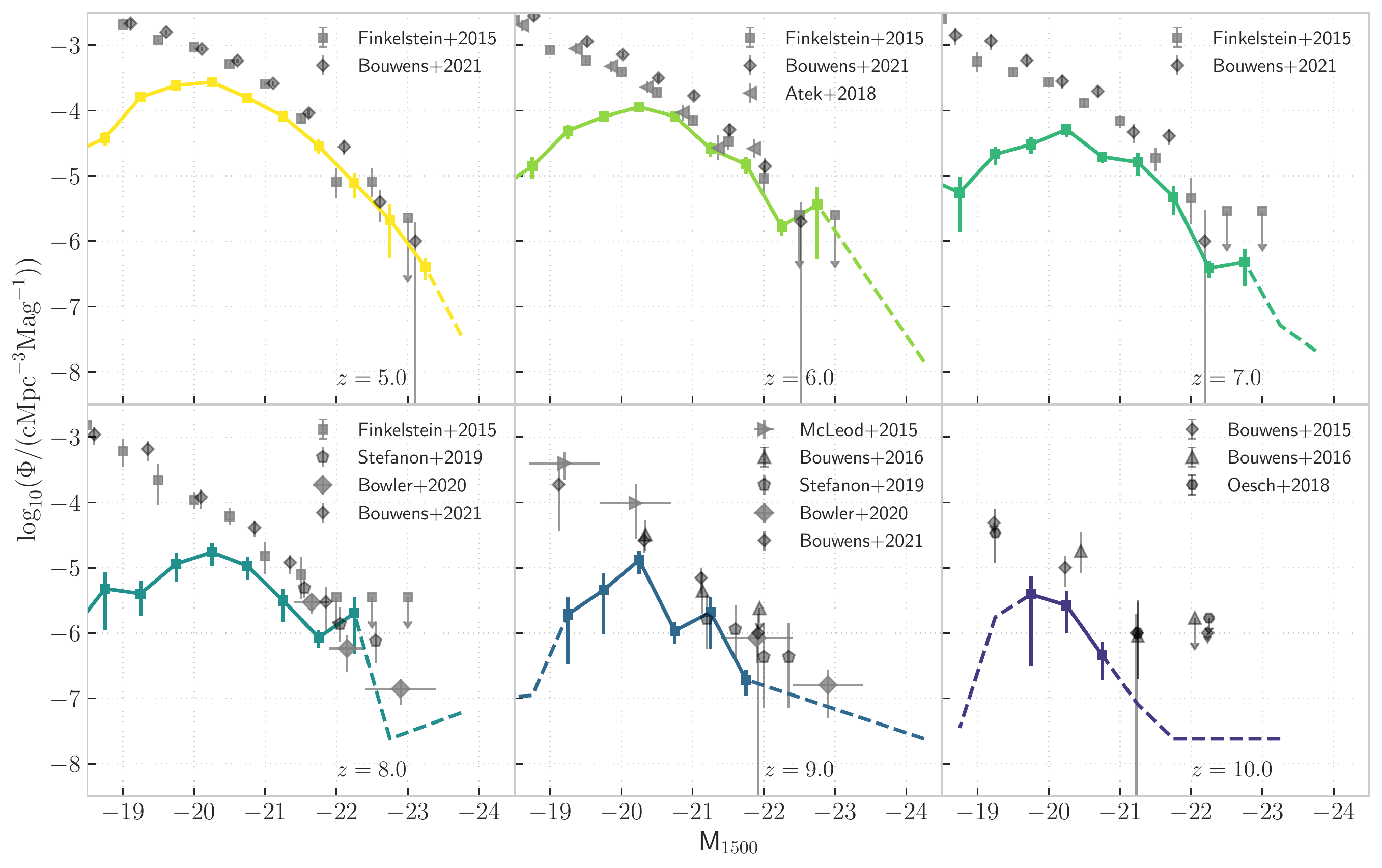}
	\caption{UV luminosity function of the \flares\ galaxies for $z\in[5,10]$. The errorbars show the Poisson $1\sigma$ uncertainties for the different bins. Bins with fewer than 5 galaxies are represented by dashed lines. The data is incomplete at the faint-end due to our galaxy selection. We also plot alongside observational data from \protect\cite{McLeod2015,Finkelstein2015,Bouwens_2016,Bouwens2017,Oesch_2018,Atek2018,Stefanon2019,Bowler2020,Bouwens2021}.\label{fig: UVLF}}
\end{figure*}

\begin{figure*}
	\centering
	\includegraphics[width=\textwidth]{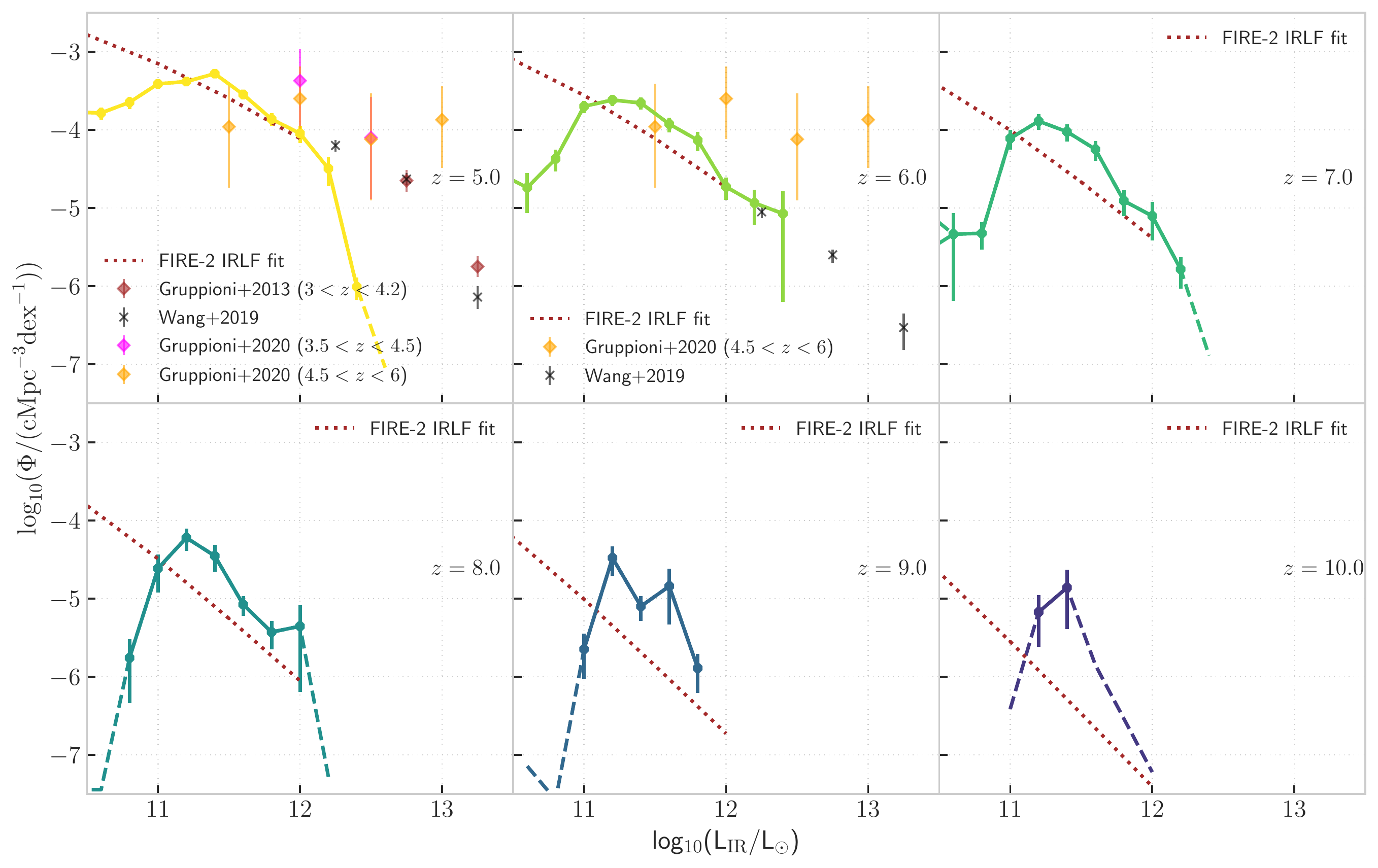}
	\caption{IR luminosity function of the \flares\ galaxies for $z\in[5,10]$. The errorbars show the Poisson $1\sigma$ uncertainties for the different bins. Bins with fewer than 5 galaxies are represented by dashed lines. The data is incomplete at the faint-end due to our galaxy selection. We plot alongside observational data from \protect\cite{Gruppioni2013,Wang2019,Gruppioni2020}. Also plotted is the IR LF fit from the \textsc{Fire-2} simulations \protect\cite[][]{Ma2019}. \label{fig: IRLF}}
\end{figure*}

\begin{figure*}
	\centering
	\includegraphics[width=\textwidth]{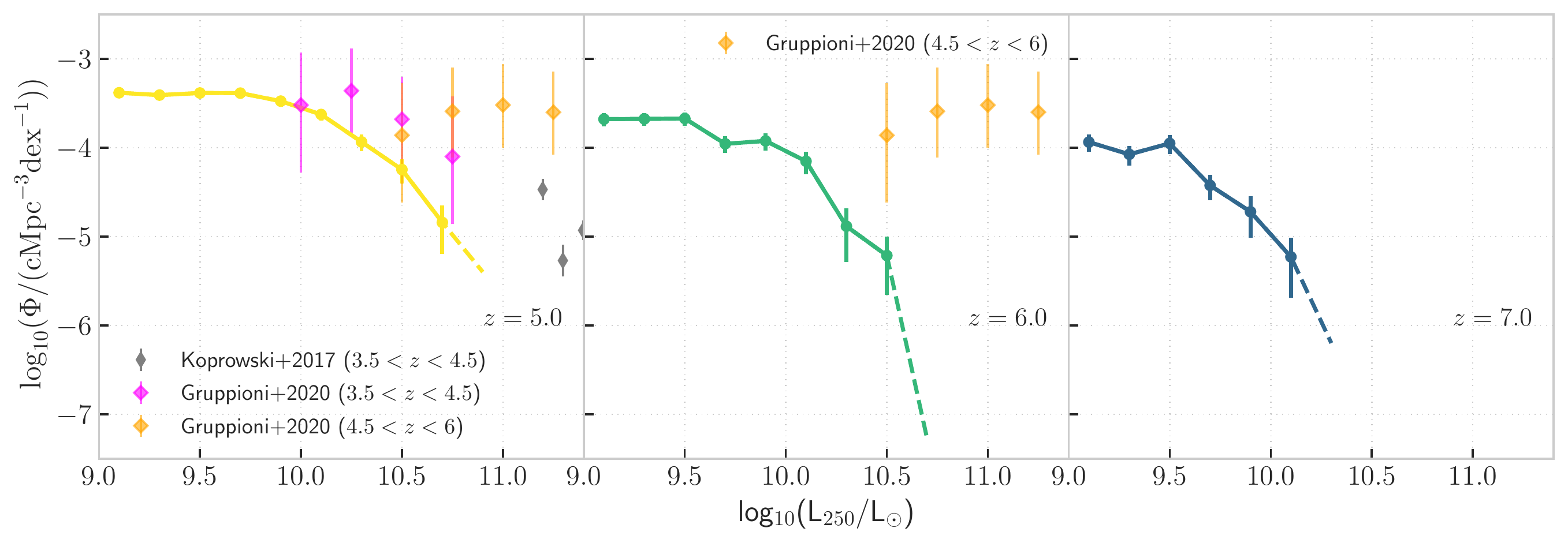}
	\caption{The rest-frame 250 \mumetre\ luminosity function of the \flares\ galaxies for $z\in[5,7]$. The errorbars show the Poisson $1\sigma$ uncertainties for the different bins. Bins with fewer than 5 galaxies are represented by dashed lines. The data is incomplete at the faint-end due to our galaxy selection. We plot alongside observational data from \protect\cite{Koprowski2017,Gruppioni2020} at similar redshift range. \label{fig: L250LF}}
\end{figure*}

In Fig.~\ref{fig: UVLF} we show the observed UV luminosity (measured at 1500\AA, plotted in magnitudes) function for the \flares\ galaxies for $z\in[5,10]$. 
This is an observational space where there is plenty of data and we compare our results to data from \cite{McLeod2015,Finkelstein2015,Bouwens_2016,Bouwens2017,Oesch_2018,Atek2018,Stefanon2019,Bowler2020,Bouwens2021}. 
The UV LF is also usually used in calibration of dust models in high-redshift theoretical studies \cite[\eg][]{Wilkins2017,Vogelsberger2020,Vijayan2021}. 
As can be seen, our model reproduces the UV LF reasonably well within the scatter seen in the observational data for M$_{1500}\lesssim-21$. The turnover at the faint-end is mainly due to our selection of well-resolved massive galaxies, whose contribution are at the bright end.
It should be noted that there is a hint of galaxy number densities being slightly lower at $z=8$ compared to the observations. There is also a similar trend at $z=10$, but is harder to draw conclusions from, since the \cite{Oesch_2018} data contains only 5 galaxies. Some of this tension can be attributed to the slightly lower normalisation ($\sim0.3$ dex) of the SFR function of the \eagle\ reference volume or \flares\ at intermediate SFR ($1 < \mathrm{SFR} < 10$ M$_{\odot}$yr$^{-1}$) as noted for high-redshift galaxies in \citet[][also see \citealt{furlong_evolution_2015}]{katsianis_evolution_2017,Lovell2021a} when compared to observed values. \cite{Vijayan2021} also showed that the unobscured SFR density of \flares\ galaxies at $z\in[5,7]$ showed slightly lower normalisation ($\sim0.2$ dex) in comparison with the unobscured value from \cite{Bouwens2020}, even though the dust model was explicitly calibrated to match the UV LF, indicating that either the star formation rates are generally lower in the simulation, or the chemical enrichment rate (and thus the derived dust content) is higher, giving rise to higher attenuation than expected in these model galaxies. 
In the future with \jwst\ we will be able to put tighter constraints on galaxy metallicities in the high-redshift regime. There is really good agreement at the high UV luminosity end at all the redshifts. 
Since the UV LF is predicted considerably well against observations (with the caveats noted) we will now try to draw meaningful conclusions from comparing against other observational spaces.

In Fig.~\ref{fig: IRLF} we show the IR luminosity functions for $z\in[5,10]$. The IR luminosity of the \flares\ galaxies are obtained by integrating the observed SED between rest-frame wavelength of $8-1000$ \mumetre. 
We also plot alongside observational data from \citet[using \textit{Herschel} data]{Gruppioni2013}; \citet[using the \textit{Herschel} catalogue generated by the Bayesian source extraction tool \textsc{xid+} in the COSMOS field]{Wang2019}; \citet[using the the ALPINE-ALMA data]{Gruppioni2020} as well as theoretical results from \textsc{Fire-2} \cite[][their IR LF fit obtained from running \skirt]{Ma2019} for similar redshifts. The \textsc{Fire-2} IR LF fit is only plotted till L$_{\mathrm{IR}}=10^{12}$ L$_{\odot}$ due to the bulk of their sample being mostly lower mass galaxies compared to \flares. It can be seen that \flares\ is in agreement with the observational data for luminosities $\lesssim10^{12}$ L$_{\odot}$ for $z=5$. There is a sharp decline in extremely bright IR galaxies in our simulation at $z=5$. The very bright end of the function is under-estimated by $\sim1$ dex (similar to the \textsc{Fire-2} IR LF fit) compared to \cite{Gruppioni2013,Gruppioni2020}, which are collated measurements within broad redshift ranges. Due to this broader redshift range, the normalisation can be higher, since lower redshifts are expected to have higher number densities. However, a difference of $\sim1$ dex is in tension with our predictions. \cite{Zavala2021} have also described the IR LF measurements in \cite{Gruppioni2020} to be representative of an overdense patch in the high-redshift Universe. This inference comes from the observational targets being highly clustered massive galaxies (log$_{10}$(M/M$_{\odot}$)$\gtrsim10.5$). This is in good agreement with the plotted IR LF of highly overdense regions in \flares\ shown in Fig.~\ref{fig: env_lir}.
In case of the \cite{Wang2019} data at $z=5$ there is a similar case of underprediction of bright IR luminous galaxies. 
Thus we are inconsistent in a regime where two independent measurements \cite[]{Gruppioni2013,Wang2019} agree, though it should also be noted that they are both obtained from the \textit{Herschel} catalogue, and are subject to uncertainties associated with the deblending techniques employed. Thus they can be ideally treated as upper limits on the IR luminosity function.

A reason for this sudden decrease is that our extreme IR-bright galaxies are biased towards the most overdense regions, having much lower contribution to the IR LF (see \S\ref{sec:res::LF::env}). Following from our argument in the UV LF section, the lower normalisation of the star formation rate function in our model at these redshifts also contributes to this lower number density \cite[also discussed in][]{McAlpine2019,Baes2020}. 
This has also been investigated at lower redshifts and has been similarly attributed to the lower star formation rate as well as the lack of `bursty' star formation in the \eagle\ model \cite[see][]{McAlpine2019}. 
However, our result is not an isolated case and has been a feature of many other cosmological and zoom simulations like \textsc{Illustris-Tng} \cite[][]{Shen2022} and \textsc{Fire-2} \cite[also plotted in Figure~\ref{fig: IRLF}]{Ma2019} at these redshifts. 
The \textsc{Simba} \cite[]{Dave2019} suite of simulations shows a higher normalisation of the SFR function than \eagle\ at high-redshift. 
In \cite{Lovell2021b}, where they post-processed \textsc{Simba} galaxies using \textsc{Powderday}, \citep{Narayanan2021} they find reasonable agreement with observationally inferred 850 \mumetre\ number counts, which they partly attribute to the higher SFRs.
So the lower star formation rate at high-redshift is a likely cause of the deficit in IR luminosities in our model. 

There are caveats that come along with physics recipes to produce higher star formation. 
For example, there is a lack of quiescent galaxies at high-redshifts in \textsc{Simba} compared to observations and \eagle, as explored in \cite{Merlin2019}. 
A fine interplay of feedback and star formation is fundamental to match the various observational results and thus provide test beds for improving the model recipes.
There has also been suggestions of changes to the initial mass function to a top heavy one in the most luminous galaxies to produce the seen higher number density of IR luminous galaxies \cite[see for \eg][]{Motte2018,Schneider2018,Zhang2018}. Any of these two scenarios would imply higher dust content from increased star-formation, thus reconciling the increase in the intrinsic emission with higher attenuation. 

The \flares\ IR LF at $z=6$ is in very good agreement with the \citet[also similar to what was seen for the IR LF of \eagle\ galaxies in that study]{Wang2019} and \textsc{Fire-2} results, while still being more than $\sim0.5$ dex lower compared to the \cite{Gruppioni2020} data for $4.5 \le z \le 6$. At $z=7$ the \flares\ IR LF shows a higher normalisation ($\sim0.3$ dex) compared to the fit from \textsc{Fire-2} in the range $10^{11}-10^{12}$ L$_{\odot}$. There are no observational constraints on the IR LF for $<10^{11}$ L$_{\odot}$ at $z>4$ yet, hence it is hard to draw conclusions from the \textsc{Fire-2} data.

In Fig.~\ref{fig: L250LF} we plot the rest-frame 250 \mumetre\ luminosity function of the \flares\ galaxies in $z\in[5,7]$. We also compare to observational data from \citet[using sub-mm/mm imaging from \textit{SCUBA-2} and \textit{ALMA}]{Koprowski2017}; \cite{Gruppioni2020}. 
In general, we see an underprediction of the rest-frame 250 \mumetre\ luminosity function. We are in agreement with the $3.5<z<4.5$ \cite{Gruppioni2020} data within the uncertainties, while at $z=6$, our results are $\sim1$ dex lower at the extreme bright end. It can also be seen that the \cite{Gruppioni2020} data in the two redshift range show no clear decline in the number density galaxies, while in our case there is a reduction in number density by $\sim0.5$ at the bright end of the function. But this is inconsistent with \cite{Koprowski2017} where they found a lower number density in the range $3.5<z<4.5$ compared to the \cite{Gruppioni2020} values. This discrepancy in the two data sets could be due to incompleteness in the sample selection associated with the \cite{Koprowski2017} data. Nevertheless, our data does not extend to the extreme luminosities that the \cite{Koprowski2017} sample covers. 

In Appendix~\ref{sec:AGN} we add the AGN bolometric luminosity to the IR luminosity for comparison to observations to gauge the effect AGN has on the IR LF. We see very small changes, not enough to reconcile an order of magnitude difference at the bright end of the IR LF. Also, in Appendix~\ref{sec:BC emission}, we look at how our results are consistent with the current setup when excluding emission from birth clouds of young stars for $z\ge8$.

\subsubsection{Environmental Dependence of IR LF}\label{sec:res::LF::env}    

\begin{figure*}
	\centering
	\includegraphics[width=\textwidth]{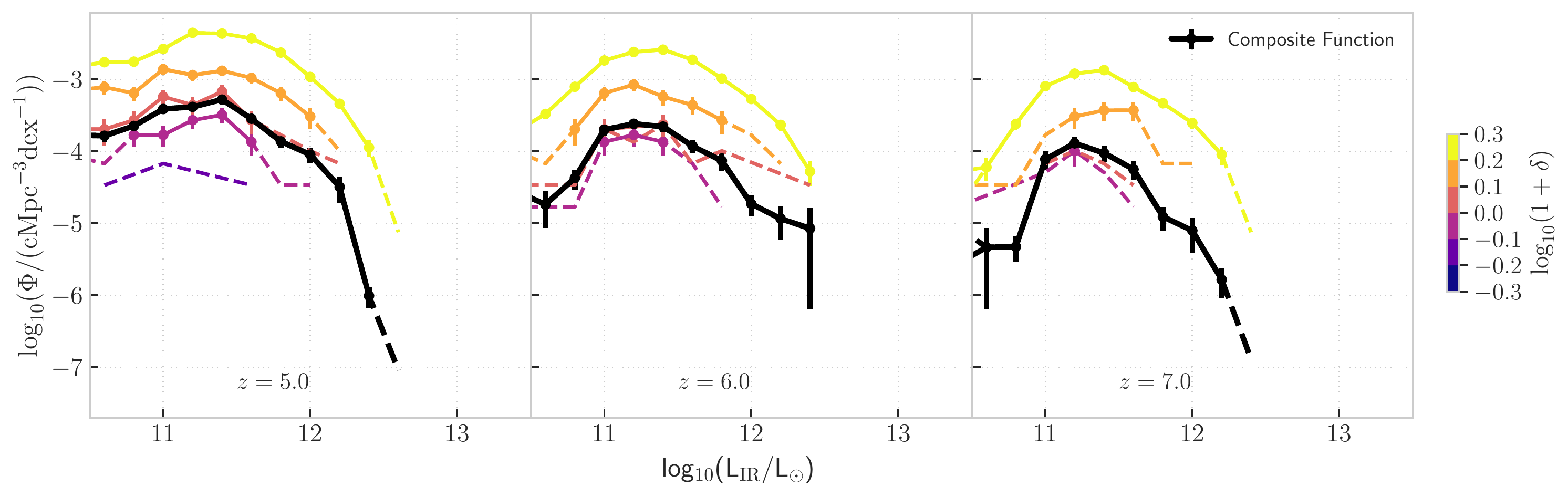}
	\caption{The \flares\ IR LF for $z \in [5,7]$ split by binned log-overdensity. Error bars denote the Poisson $1\sigma$ uncertainties for each bin from the simulated number counts. The composite distribution function is plotted as black solid line. \label{fig: env_lir}}
\end{figure*}

In Fig.~\ref{fig: env_lir}, we show the IR LF in different matter overdensity bins for $z\in[5,7]$. 
The composite function sits within the boundary of the positive and negative overdensity bins as expected. 
The plot shows that there is an increase in the number densities of IR luminous galaxies with increasing overdensity. 
The smallest overdensity bin is not visible in the plot since it is below the plotted IR luminosity range. At $z=5$, there is an increase of $\sim1.5$ dex in number density of galaxies with IR luminosity of $\sim 10^{11}$ L$_{\odot}$. This is very similar across the rest of the plotted redshift range. 
At $z=5$, only the most overdense regions contribute to the very bright end of the IR LF, which results in the rapid fall in the number densities seen in the composite function.

\subsection{IRX-$\beta$}\label{sec:res::IRXbeta}

\begin{figure*}
	\centering
	\includegraphics[width=\textwidth]{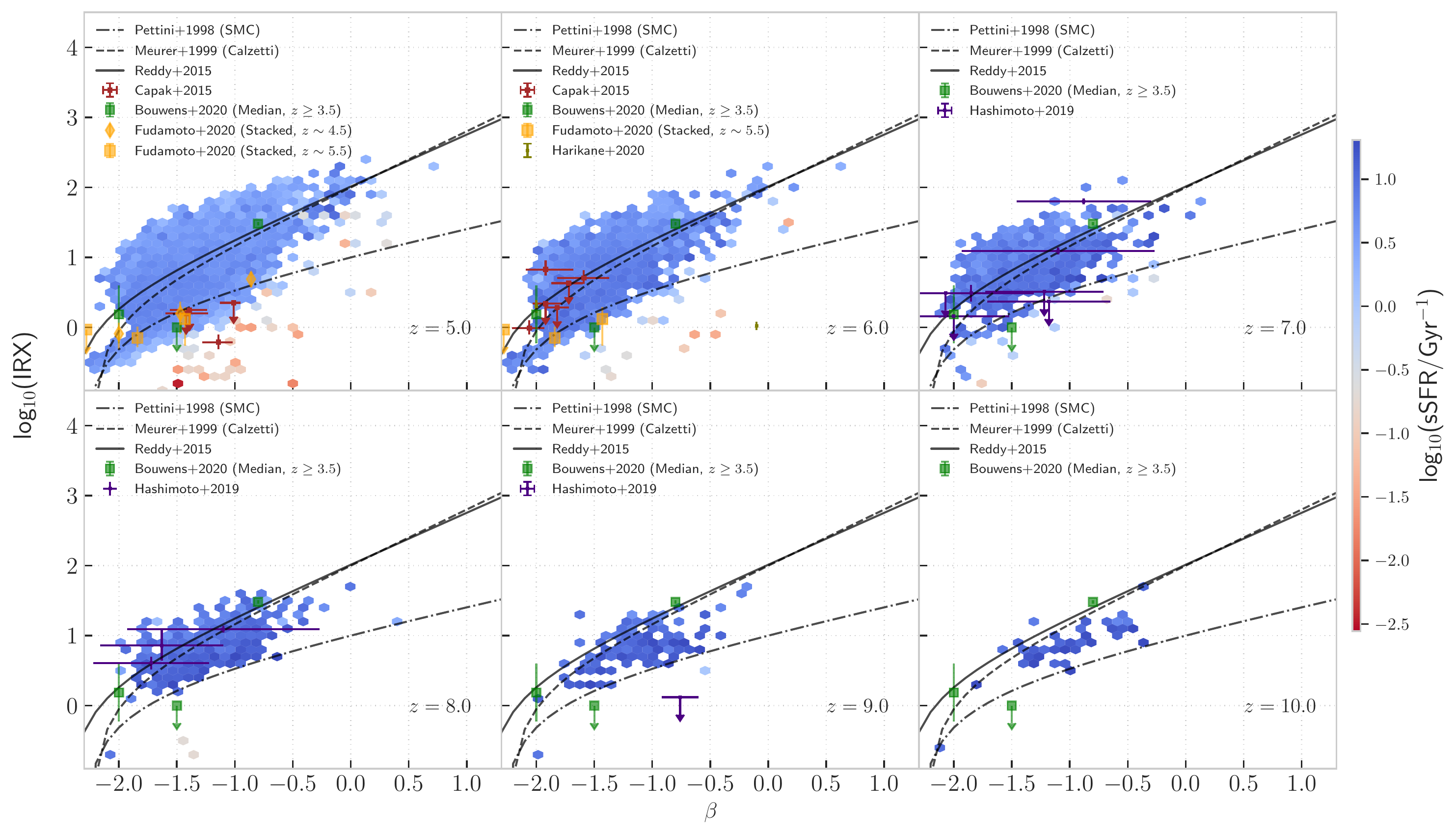}
	\caption{IRX-$\beta$ distribution of the \flares\ galaxies for $z\in[5,10]$. The hexbins are coloured by the median specific star formation rate. We also show individual observational data at similar redshift from \protect\citet[updated $\beta$ and IRX values from \protect\citealt{Barisic2017} and \protect\cite{Casey2018b} respectively]{Capak2015}, \protect\citet[also with observations collated from \protect\citealt{Ouchi2013,Ota2014,Inoue2016,Knudsen2017,Laporte2017,Hashimoto2018,Marrone2018,Smit2018,Tamura2019}]{Hashimoto2019}, \protect\cite{Harikane2020}. We also show stacked results from \protect\cite{Fudamoto2020,Bouwens2020}. Also shown is the empirical relation for dust screen models for SMC \protect\cite[][]{Pettini1998} and Calzetti \protect\cite[][]{Meurer1999}, and from \protect\cite{Reddy2015}.\label{fig: IRXbeta}}
\end{figure*}

\begin{figure*}
	\centering
	\includegraphics[width=\textwidth]{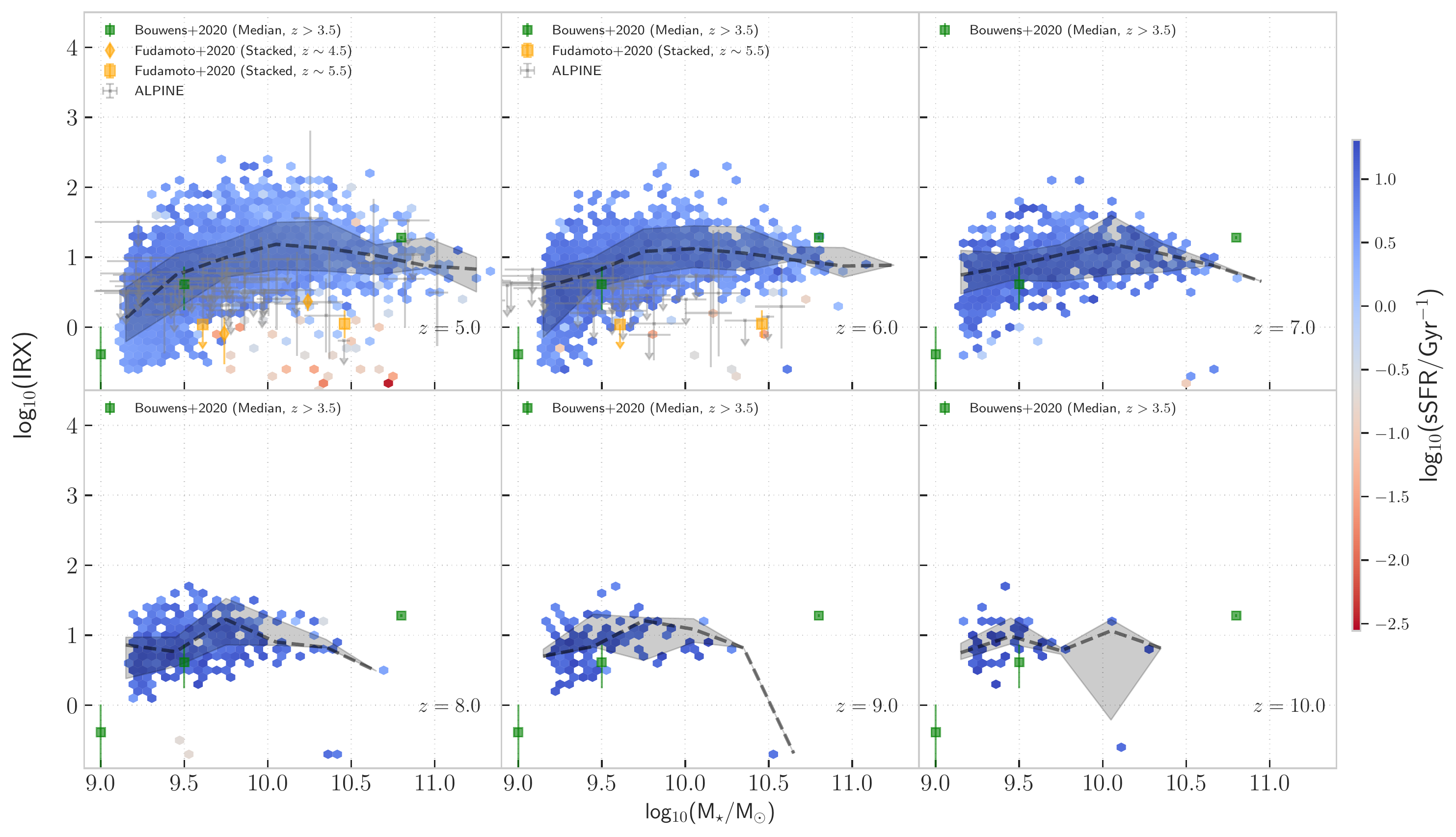}
	\caption{IRX-stellar mass distribution of the \flares\ galaxies for $z\in[5,10]$. The hexbins are coloured by the median specific star formation rate. The dashed line and the shaded region denote the weighted median and the 16-84 percentile spread of the data. We plot observational data from the publicly available ALPINE collaboration data \protect\cite[][]{LeFevre2020,Bethermin2020,Faisst2020a}. Also shown is the stacked and median relation from \protect\cite{Fudamoto2020} and \protect\cite{Bouwens2020} respectively. \label{fig: IRXmstar}}
\end{figure*}

In this section we will look at the infrared excess (IRX) of the galaxies in the \flare\ simulation. 
The infrared excess is defined as the ratio of the total infrared luminosity ($L_{\mathrm{IR}}$) over the UV luminosity ($L_{\mathrm{UV}}$), and can be derived as follows
\begin{gather}\label{eq: IRX}
	\mathrm{IRX} = \frac{L_{\mathrm{IR}}}{L_{\mathrm{UV}}} \simeq \frac{\int_{8\mumetre}^{1000\mumetre}L_{\lambda}\mathrm{d}\lambda}{L^{'}_{\mathrm{1500}}},
\end{gather} 
where $L^{'}_{\mathrm{1500}}=L_{\mathrm{1500}}\times1500$ \AA, with L$_{\mathrm{1500}}$ being the far-UV luminosity calculated at $1500$ \AA. 
The UV-continuum slope, $\beta$ is defined such that $f_{\lambda}\propto\lambda^{\beta}$ or alternatively $f_{\nu}\propto\lambda^{\beta+2}$, for $\lambda$ in the rest-frame UV range. 
We measure $\beta$ using the following prescription,
\begin{gather}\label{eq: beta}
	\beta = \frac{\mathrm{log}_{10}(L_{\mathrm{1500}}/L_{\mathrm{1500}})}{\mathrm{log}_{10}(1500/2500)} - 2,
\end{gather}
where $L_{\mathrm{1500}}$ and $L_{\mathrm{2500}}$ are the far-UV and near-UV luminosity, respectively.

The IRX-$\beta$ relation has been explored in numerous theoretical \cite[\eg][]{Popping2017,Safarzadeh2017,Ma2019,Trcka2020,Schulz2020,Liang2021} and observational studies \cite[\eg][]{Reddy2015,Wang_2018,Fudamoto2020,Bouwens2020} in low- and high-redshift galaxies. Some studies have provided empirical relations for this plane assuming a dust screen model. These relations are expected to arise from the simple assumption that with increasing dust attenuation the UV-continuum slope becomes redder with the $L_{\mathrm{IR}}$ to $L_{\mathrm{UV}}$ ratio increasing. 
The assumption of different dust attenuation curves will determine the trajectory of this relation. 
However, in galaxies one would expect different attenuation for young and old stars, as well as different dust distribution across the galaxy that can provide large variation to the relationship \cite[\eg][]{Narayanan2018,Schulz2020}.

The empirical relation between IRX and $\beta$ is widely used to correct for the amount of dust-obscured star formation within galaxies at high-redshift. 
This is based on the assumption that high-redshift galaxies follow the same relation as their local analogues. 
However, some high-redshift galaxies seem to have smaller IRX than their local analogues \cite[\eg][]{Capak2015,Fudamoto2020}. This has been largely attributed to the low dust temperatures adopted in modelling the observational data \cite[\eg][by changing the usual assumption of dust temperatures of $35-45$ K to $45-50$ K]{Sommovigo2020,Bouwens2020}. 
Another cause of concern in using this relation at high redshift is the observed spatial offset between the UV and IR emission \cite[\eg][]{Bowler2018}, possibly due to the birth cloud dispersal time in galaxies \cite[][]{Sommovigo2020}. 

We look at the variation of the IRX-$\beta$ relation across $z\in[5,10]$ in Fig.~\ref{fig: IRXbeta}. 
The plane is represented by hexbins which are coloured by their median sSFR values. 
Due to our simulations containing a large selection of extreme overdensities, there is an overabundance of IR luminous dusty galaxies in our data. We also plot observational data at similar redshifts from \citet[with updated $\beta$ values from \citealt{Barisic2017}, the IR luminosity was obtained using a power-law + MBB functional form (see equation~\ref{eq: GP-MBB}) such that the obtained SED mirrored the observed scatter in $L_{\mathrm{IR}}$ - $\lambda_{\mathrm{peak}}$ of their sample, then select the SEDs which produced the observed flux density at 1.2 mm and generate a probability density distribution in $L_{\mathrm{IR}}$]{Capak2015}, \citet[also with observations collated from \citealt{Ouchi2013,Ota2014,Inoue2016,Knudsen2017,Laporte2017,Hashimoto2018,Marrone2018,Smit2018,Tamura2019}, all obtained using optically-thin MBB function with $T_{\mathrm{d}}=50$ K and $\beta=1.5$, see equation~\ref{eq: MBB-OT}]{Hashimoto2019}, \citet[obtained by fitting observed fluxes using optically-thin MBB with $\beta=1.6$ and varying the IR luminosity and dust temperature]{Harikane2020} as well as the stacked and median results from \citet[estimated using the conversion factor from $158$ \mumetre\ to $L_{\mathrm{IR}}$ presented in \cite{Bethermin2020} using stacking of sources; caveats of the conversion being that it could be largely inaccurate for outliers with extreme dusty SEDs as well as stacking favouring brighter sources]{Fudamoto2020} and \citet[using optically-thin MBB with $\beta=1.6$ and redshift evolution of dust temperature based on their equation~1]{Bouwens2020}. Also shown is the empirical IRX-$\beta$ relations from \citet[SMC]{Pettini1998}, \citet[Calzetti]{Meurer1999} and \cite{Reddy2015}. 

From the figure we can see that IRX-$\beta$ relation of the \flares\ galaxies agree well with plotted observational values. There are a few exceptions in case of data with high-$\beta$ and low-IRX values (low dust content and older stellar populations) found for a few galaxies in the \cite{Hashimoto2019} and \cite{Harikane2020} sample. This could be a drawback of our implemented model that does not fully capture the diverse star-dust geometries of galaxies in these observations. These galaxies having high-$\beta$ and low-IRX imply that they are moving away from the main-sequence relation towards quiescence. The \flares\ sample could be inefficient in producing such galaxies. The quiescent galaxy population in \flares\ will be probed in a future work where their number densities will also be explored. A similar dearth of high-$\beta$, low-IRX galaxies is seen in \cite{Ma2019}. However, their probed galaxy stellar mass range is lower than ours and thus it could also be due to the fact that there are not any low-sSFR high-mass galaxies, that are moving into the quiescent regime.

At high-redshifts ($z>4$), no clear picture has emerged on what kind of attenuation relation galaxies follow, whether or not they are consistent with the local relation, following a starburst or Calzetti--like attenuation law, or a steeper one like the SMC. As can be seen in Fig.~\ref{fig: IRXbeta}, the massive galaxies in \flares\ predominantly follow the Calzetti or \cite{Reddy2015} like relation, with a small proportion of galaxies following the SMC relation at $z\in[5,8]$, even though we adopted the SMC grain distribution. 
We also note that the galaxies which drop below the canonical relations are the ones that exhibit very low sSFR values, or with older stellar populations, in agreement with theoretical studies like \cite{Narayanan2018}. 
At higher redshifts ($z\ge8$) there is a hint of some transition away from the Calzetti relation towards the SMC one in \flares, which we fail to properly capture due to the limited mass resolution of our simulation. This leads to the lack of well resolved low-mass galaxies in our sample to populate this space. 

The reason for the majority of \flares\ galaxies following the local starburst relation is mostly because these high-redshift galaxies are intrinsically bluer with higher sSFR compared to their low-redshift counterparts. The inhomogeneous nature of dust distribution at high redshift will also have an effect, with the $\beta$ values being dominated by unobscured young stars while the IRX is dominated by dust emission near the highly obscured dust patches. With our selection of the most massive galaxies, this is a very likely outcome due to their higher dust content.
\cite{Narayanan2018}, using cosmological zoom simulations run using \textsc{Gizmo} \cite[]{Hopkins2015}, also attributed the major drivers of the observed deviations from the canonical relation to older stellar populations, complex star-dust geometries and variations in dust extinction curves. 
A similar result was also seen at high-redshift in \cite{Schulz2020}, studying the IRX-$\beta$ relation in \textsc{Illustris-Tng} galaxies at $z=0-4$. 
They concluded that the seen deviations could be best described in terms of sSFRs driving the shift in $\beta$ with the star formation efficiency possibly being a good indicator of the variations in star-dust geometry. 
\cite{Liang2021} explored in detail the secondary dependencies of the IRX-$\beta$ relation, concluding that the main driver of the scatter is the variations in the intrinsic UV spectral slope and thus the age of the underlying stellar population. 
Thus, due to large degeneracies among sSFRs or ages, star-dust geometry as well as dust compositions, it would be hard to pin-point a global track in the IRX-$\beta$ relation for galaxies at these redshifts for different stellar masses. 

In Fig.~\ref{fig: IRXmstar} we look at the relationship between the galaxy stellar mass and IRX. We also plot observational results from the publicly available dataset of the ALPINE collaboration \cite[][values obtained from SED fitting]{LeFevre2020,Bethermin2020,Faisst2020a} as well as the stacked and median results from \cite{Fudamoto2020} and \cite{Bouwens2020} respectively. Also shown is the weighted median and the 16-84th percentile variation for the sample of galaxies. This plane provides further insights into what we saw in Fig.~\ref{fig: IRXbeta}. We can see that the galaxies following the SMC relation have stellar masses of $\sim10^9$ M$_{\odot}$. These galaxies are the ones in the process of transitioning to the higher attenuation relation from rapid dust enrichment (see figs~9 and 10 in \citealt{Vijayan2021} where we also see a rapid rise in the UV attenuation $\sim10^9$ M$_{\odot}$ in stellar mass). We also see that there is a general trend towards high median IRX values with higher stellar masses, plateauing or slightly dropping at the highest masses. If we extrapolate the median to lower masses, it would lead to lower IRX values. This would indicate that the region near the SMC relation would be occupied by the lower mass galaxies. This is in agreement with what was seen in \cite{Ma2019} using the FIRE-2 simulation, where the majority of the galaxies below a stellar mass of $10^9$ M$_{\odot}$ follow the SMC relation. 

The trend at the massive end ($\sim10^{10}$ M$_{\odot}$), which is clear at $5\le z\le8$, points towards an increase in the UV luminosity not being reflected to the same extent in the IR luminosity. This points towards decreasing dust attenuation in the most massive galaxies. This was also seen in the LOS dust attenuation model applied on the \flares\ galaxies in \citet[][see fig.~10]{Vijayan2021}. The observed UV LF being better fit by double power-law at these high-redshift \cite[\eg][]{Bowler2014,Bowler2020,Shibuya2022} also points towards decreasing dust attenuation at the bright and massive end, that can contribute to this decline. However, studies such as \cite{Ferrara2017} posit that some of the IRX deficit galaxies could have dust embedded in large gas reservoirs, thus not contributing to an increase in the IR luminosity. 

On comparing to the observational data, the median values from \citet[note that the highest mass bin has just one galaxy]{Bouwens2020} are a good match. In case of the ALPINE data as well as the stacked results from \citet[which also uses ALPINE data]{Fudamoto2020}, our median relation is higher than their dataset. However, this can be explained by the UV selection of the galaxies observed in the survey, which can miss the dustier systems. 

There is no noticeable evolution of IRX-stellar mass relation with redshift. The normalisation of the median value does not show any redshift evolution.

\subsection{Dust temperatures}\label{sec:res::Td}
In this section we will look at the dust temperatures of the galaxies in \flares. There are different definitions of the dust temperature, both observational and theoretical. Here we will look at the SED--derived and peak wavelength temperature, which are both measures of the light-weighted dust temperatures, but can be considerably different depending on the functional forms being used \cite[see e.g.][]{Casey2012}. 
These measures are very different to the mass-weighted temperature, a measure of mostly the cold dust content of galaxies, which is expected to be largely independent of redshift and galaxy properties as well as significantly lower than the light-weighted temperatures \cite[][]{Liang2019,Sommovigo2020}.

\subsubsection{T$_{\mathrm{peak}}$}
The peak dust temperature ($T_{\mathrm{peak}}$) can be obtained from the Wein displacement law from the rest-frame wavelength at which the infrared flux density peaks ($\lambda_{\mathrm{peak}}$). 
This is defined as 
\begin{gather}
T_{\mathrm{peak}} = \frac{2.898\times 10^3}{\lambda_{\mathrm{peak}}} \mumetre\,\mathrm{K},
\end{gather}
which follows the relation for a true blackbody (that has a dust emissivity index $\beta$ of 2). This measure has been used in many observational studies to understand the evolution of luminosity-weighted dust temperature across redshifts and galaxy properties \cite[\eg][]{Casey2018,Schreiber2018,Burnham2021}. 
It suffers less from model dependent biases compared to other dust temperature values. 
It should also be noted that $T_{\mathrm{peak}}$ is only a proxy for $\lambda_{\mathrm{peak}}$, and the choice of the normalisation is to compare with other theoretical and observational studies.

\begin{figure*}
	\centering
	\includegraphics[width=\textwidth]{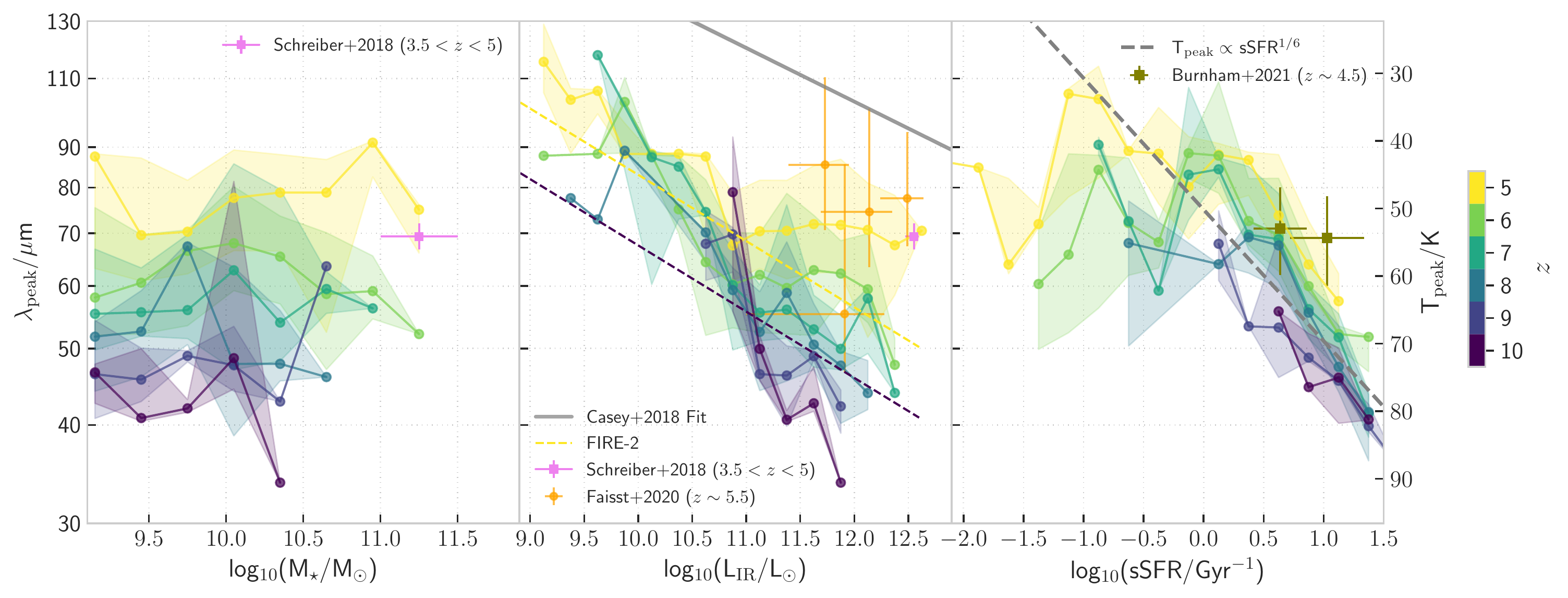}
	\caption{We show the variation of $\lambda_{\mathrm{peak}}$ (corresponding $T_{\mathrm{peak}}$ values is shown on the right y-axis) with various galaxy properties like the galaxy stellar mass (left panel), total infrared luminosity (middle panel) and the specific star formation rate (sSFR, right panel) for $z\in[5,10]$. The weighted median (solid line) and the 16-84 percentile (shaded region) variation is plotted fo the different redshifts which are denoted by the different colours as per the shown colourbar. We also show observational data from \protect\citet[median data in $3.5<z<5$ for the stellar mass range of $10^{11}-10^{11.5}$ M$_{\odot}$]{Schreiber2018} in the stellar mass-$\lambda_{\mathrm{peak}}$ plane and the measurements from \protect\citet[for $z\sim5.5$]{Faisst2020b} in the IR luminosity-$\lambda_{\mathrm{peak}}$ plane. The fits obtained from observations of $z<5$ galaxies from \protect\citet[][grey solid line]{Casey2018} and \textsc{Fire-2} \protect\cite[for $z=5,10$, with same colour as the \flares\ median lines, but dashed]{Ma2019} simulations are also shown. We also plot a T$_{\mathrm{peak}}\propto$ sSFR$^{1/6}$ relation to compare to our median relations in the right panel. \label{fig: lampeakprop}}
\end{figure*}

In Fig.~\ref{fig: lampeakprop}, we show how the weighted median as well as the 16-84th percentile (shaded region) peak of the IR emission varies with various galaxy properties like the stellar mass (left panel), IR luminosity (middle panel) and the specific SFR (sSFR, right panel) for $z\in[5,10]$. 
In case of the variation in $\lambda_{\mathrm{peak}}$ with galaxy stellar mass, we do not find any significant trend in the ranges we are considering. There is a general decrease (increase) in the $\lambda_{\mathrm{peak}}$ ($T_{\mathrm{peak}}$) value with increasing redshift. 
The observational data from \cite{Schreiber2018}, which shows the median data in $3.5<z<5$ in the stellar mass range $10^{11}-10^{11.5}$ M$_{\odot}$, is in agreement with our data at $z=5$. 

In a similar vein, the middle panel of Fig.~\ref{fig: lampeakprop} shows the variation of $\lambda_{\mathrm{peak}}$ with the IR luminosity for different redshifts. We also plot along with it observational data from \citet[same range as before]{Schreiber2018} and \citet[for galaxies at $z\sim5.5$]{Faisst2020b} as well as the redshift independent relation presented in \cite{Casey2018} for $z<5$ samples. Also shown is the redshift dependent fit to the \textsc{Fire-2} \cite[post-processed with \skirt]{Ma2019} for $z=5,10$. 
The data from \cite{Strandet2016,Faisst2020b} match well with our constraints at $z=5$, while the relation from \cite{Casey2018} is well above our values as well as the high-redshift observations. 
The \flares\ galaxies agree well with the $z=5$ fit from \textsc{Fire-2}, but the slope of the relation at higher redshifts ($z\ge7$) is steeper in \flares{}. There is a trend of lower (higher) values of $\lambda_{\mathrm{peak}}$ ($T_{\mathrm{peak}}$) with increasing IR luminosity similar to the what is seen in \cite{Ma2019} and \citet[with the \skirt, post-processed \textsc{Illustris-Tng} galaxies]{Shen2022}. However, by $\sim10^{11}$L$_{\odot}$ we see a flattening in this relation similar to what was found in \cite{Shen2022} at $z=4,6$, with a higher normalisation than the one here. As posited there as well as in other studies \cite[]{Jin2019}, this trend could be due to the increasing optical depth in the most luminous galaxies, hiding the warm dust associated with star-forming compact regions, making the contribution to the dust temperature minimal. 
At the high IR luminosity end, there is a strong evolution towards lower (higher) $\lambda_{\mathrm{peak}}$ ($T_{\mathrm{peak}}$) values with redshift, showing that \flares\ also prefers an evolving relation similar to results in \cite{Ma2019,Shen2022}.  

The right panel of Fig.~\ref{fig: lampeakprop} shows the variation of $\lambda_{\mathrm{peak}}$ with the sSFR (SFR calculated using stars born in the last 10 Myr) for different redshifts. We have also over-plotted two galaxies at $z\sim4.5$ from \cite{Burnham2021} which are in agreement with our $z=5$ relation within the scatter. We see a very tight relation for the \flares\ galaxies at the high sSFR (sSFR/Gyr$^{-1}\gtrsim0$) end. 
This has been observed in studies like \cite{Magnelli2014,Ma2019}. This strong correlation can be understood by looking at the following relation for an isothermal modified blackbody \cite[see][]{Hayward2011},
\begin{gather}
	\mathrm{T}_{\mathrm{dust}}\propto\bigg(\frac{L_{\mathrm{IR}}}{M_{\mathrm{dust}}}\bigg)^{1/6} \;\;(\text{for $\beta=2$}).
\end{gather}
SEDs can be qualitatively described by such a form. 
For main sequence galaxies, $L_{\mathrm{IR}}$/$M_{\mathrm{dust}}$ has been found to be proportional to the sSFR \cite[\eg][]{Magdis2012,Magnelli2014,Ma2019}, implying an inverse correlation with $\lambda_{\mathrm{peak}}$. 
We show on the figure that this matches our results well at high sSFR (the $T_{\mathrm{peak}}\propto$ sSFR$^{1/6}$ dashed line). This relation can also be used to understand the increasing dust temperatures with redshift (explicit redshift evolution is shown in Fig.~\ref{fig: Tdustevo}). \cite{Lovell2021a} has already shown that there is a systematic increase in the normalisation of the sSFR of the \flares\ galaxies at constant stellar mass. This would imply that the redshift dependence of the dust temperature can be attributed to the increasing sSFR. We explore the evolution of $\lambda_{\mathrm{peak}}$ with different stages of galaxy star-formation activity in Appendix \ref{sec: tpeak and MS}.   


\subsubsection{T$_{\mathrm{SED}}$}
To obtain the SED dust temperature we follow \cite{Casey2012} by parameterising our galaxy SEDs using the sum of a single modified-blackbody and a mid-infrared powerlaw. The addition of the powerlaw to the functional form provides a better fit to the mid-infrared which is dominated by warm dust. Using this prescription, the luminosity at a rest-frame frequency $\nu$ can be written as
\begin{gather}\label{eq: GP-MBB}
	L_{\nu}(\nu) = N_{\mathrm{bb}}f(\nu,\beta,T_{\mathrm{SED}}) + N_{\mathrm{pl}}\,(\nu_c/\nu)^{-\alpha}\,e^{-(\nu_c/\nu)^{2}}.
\end{gather}
Here,
\begin{gather}
	f(\nu,\beta,\mathrm{T}) = \frac{(1-\exp(-(\nu/\nu_1)^{\beta}))}{\exp(h\nu/(k_b T))-1}\nu^3\,,\\
	N_{\mathrm{pl}} = N_{\mathrm{bb}}f(\nu_c,\beta,T_{\mathrm{SED}}),	
\end{gather}
where $h$ and $k_b$ are the Planck's constant and the Boltzmann constant, respectively. The free parameters in the form are $N_{\mathrm{pl}}$ (normalisation factor), $\beta$ (emissivity index), $T_{\mathrm{SED}}$ (SED dust temperature), $\nu_1$ (frequency where optical depth is unity, usually taken as $\sim100$ \mumetre\ or 3 THz in high-redshift studies) and $\alpha$ (mid-IR power-law slope). We adopt the same parameterisation of $\nu_{c}$ as given in \cite{Casey2012} (or $\lambda_{c}$ there). We use \textsc{lmfit} \cite[][]{lmfit}, a non-Linear least-squares minimization and curve-fitting package in \textsc{python}, to fit this parametric form to the \textsc{skirt} SEDs and obtain $T_{\mathrm{SED}}$. 
In the fitting, we impose the criteria that the dust temperature is higher than the CMB temperature at that redshift. It should also be noted that there is degeneracy between the dust temperature and the emissivity; a higher temperature can compensate for a lower value of the emissivity, and vice-versa. Thus when comparing to observations there is considerable maneuverability when choosing the values of $T_{\mathrm{SED}}$ and $\beta$, and thus deviations or agreement with the values can also be achieved based on the ranges being probed. In our case we see that in general the SED temperature increases while the emissivity decreases with increasing redshift when they are both kept as free parameters.

The Rayleigh-Jeans (RJ) part of the SED can also be approximated by a generalised modified-blackbody function in the optically thin case \cite[equation~2 in][]{Casey2012} by
\begin{gather}\label{eq: MBB-OT}
	L_{\nu}(\nu) = A\frac{\nu^{3+\beta}}{\exp(h\nu/(k_b T_{\mathrm{SED,RJ}}))-1}\,\,,
\end{gather}
where $A$ is a normalisation constant and the rest of the terms are defined as before. 
In this case the free parameters are $A$, $\beta$ (parameter search is confined to the range $1.5-2.5$ to be consistent with high-redshift observational works referenced here) and $T_{\mathrm{SED,RJ}}$. We refer to the SED dust temperature obtained from this functional form as $T_{\mathrm{SED,RJ}}$. Similar to the fitting function in equation~(\ref{eq: GP-MBB}), a similar degeneracy exists here as well. 
This form is usually used by observational studies to derive the total infrared luminosity of galaxies \cite[\eg][]{Knudsen2017,Hashimoto2019,Bouwens2020}. 
In many cases where there is only a single detection in the dust-continuum, $\beta$ and $T_{\mathrm{SED,RJ}}$ are kept constant and a fit for the normalisation is obtained. 
It should also be noted that the SED dust temperature obtained from this form closely matches with the galaxy peak dust temperatures, while $T_{\mathrm{SED}}$ is typically higher than the peak dust temperature \cite[see fig.~2 in][]{Casey2012}. From our analysis we also see that this form can lead to an overprediction of the obtained total infrared luminosity (median deviation of $\sim+17$ per cent at $z=5$ and lowering to $\lesssim1$ per cent by $z=10$) compared to results using equation~(\ref{eq: GP-MBB}) or the true SED. 
This is seen when we use the full range of the dust SED. We have also tried to constrain our fits by only using wavelength ranges in the RJ tail. In this case, most fits underpredicted the total IR luminosity. 
Thus it is important to have some constraints at wavelengths short of the RJ tail to produce reliable SED temperature estimates that can retrieve the total IR luminosity. 

\subsubsection{Redshift evolution of dust temperatures}
\begin{figure}
	\centering
	\includegraphics[width=\columnwidth]{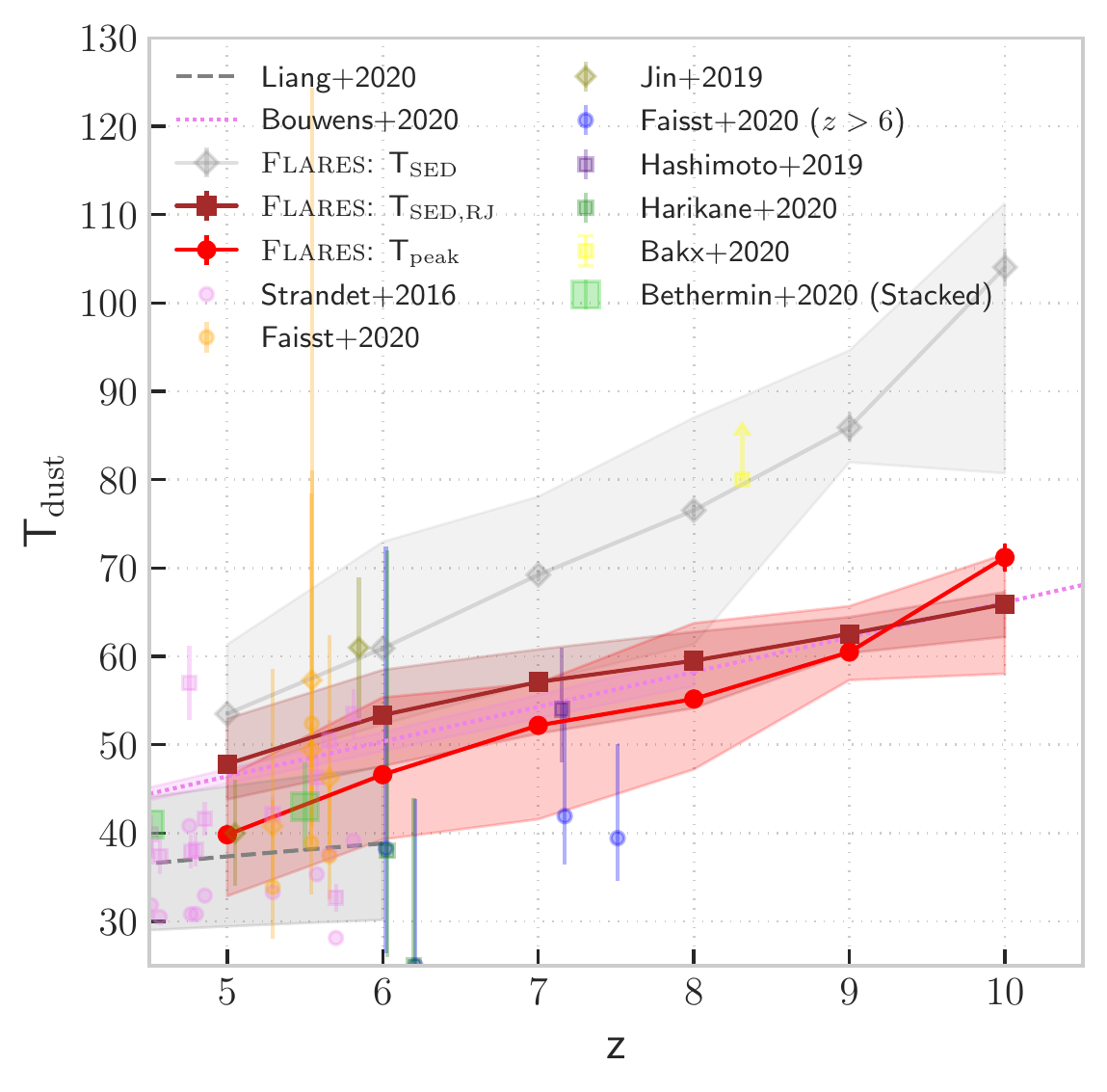}
	\caption{We show the evolution of the peak dust temperature ($T_{\mathrm{peak}}$, red circles) and the SED dust temperature (brown squares for $T_{\mathrm{SED,RJ}}$, grey diamonds for $T_{\mathrm{SED}}$) from the \flare\ simulation. The markers indicate the weighted median, the 16-84 percentile spread and the error on the median (the errorbars, negligible due to the high number counts) at $z\in[5,10]$. Observational data from studies at high redshift (circle for $T_{\mathrm{peak}}$, diamond for $T_{\mathrm{SED}}$ and square for $T_{\mathrm{SED,RJ}}$ values); included are data from \protect\citet[both $T_{\mathrm{peak}}$ and $T_{\mathrm{SED,RJ}}$]{Strandet2016}, \protect\citet[$T_{\mathrm{SED}}$ calculated from optically thick MBB]{Jin2019}, \protect\citet[both $T_{\mathrm{peak}}$ and $T_{\mathrm{SED}}$, also included are the remeasured $T_{\mathrm{peak}}$ values from \protect\citealt{Knudsen2017,Hashimoto2019}]{Faisst2020b}, \protect\citet[$T_{\mathrm{SED,RJ}}$]{Hashimoto2019}, \protect\citet[$T_{\mathrm{SED,RJ}}$]{Harikane2020} \protect\citet[$T_{\mathrm{SED,RJ}}$ obtained from stacked galaxies with SFR $\ge10\ \mathrm{M}_{\odot}$/yr]{Bethermin2020} and \protect\citet[$T_{\mathrm{SED,RJ}}$, lower limit]{Bakx2020} are plotted. We also show the fit functions for $T_{\mathrm{peak}}$ from \protect\citet[dashed line, with the shaded region showing the spread]{Liang2019} and for $T_{\mathrm{SED,RJ}}$ from \protect\citet[dotted line]{Bouwens2020} (see text for more details).\label{fig: Tdustevo}}
\end{figure}

In Fig.~\ref{fig: Tdustevo} we show the weighted median evolution of the different dust temperatures for the \flares\ galaxies (red circles for $T_{\mathrm{peak}}$, grey squares for $T_{\mathrm{SED}}$ and brown squares for $T_{\mathrm{SED,RJ}}$).
We also show the 16-84$^{\mathrm{th}}$ percentile spread of the values as well as the error on the median. 
Overplotted are several dust temperatures from observations of high-redshift galaxies (circle for $T_{\mathrm{peak}}$ and square for $T_{\mathrm{SED}}$ or $T_{\mathrm{SED,RJ}}$ values, which is explicitly stated in the figure caption and the text) as well as fit functions for $T_{\mathrm{peak}}$ and $T_{\mathrm{SED,RJ}}$ from \cite{Liang2019} (simulations) and \cite{Bouwens2020} (observations) respectively.

The figure clearly indicates that the median values of the dust temperatures consistently increase towards higher redshift, as expected since higher redshift galaxies are intensely star-forming (higher sSFR), which leads to higher UV emission resulting in warmer dust. 
There is also a spread of $\sim10$ K in all the temperature values across the redshift range. 
$T_{\mathrm{peak}}$ increases from a median value of $\sim40$ K at $z=5$ to $\sim70$ K at $z=10$, which is higher than the increase in the CMB temperature across this redshift. Using our sample of massive galaxies, we fit a linear relation to the median redshift evolution of $T_{\mathrm{peak}}$ and obtain
\begin{gather}\label{eq: Tpeakfit}
	T_{\mathrm{peak}}/\mathrm{K} = (40.03\pm0.15) + (5.77\pm0.14)(z-5).
\end{gather}
The relation has a higher slope compared to the one in \cite{Schreiber2018} (slope of $4.60\pm0.35$) obtained for an observational sample using stacked SEDs of main-sequence galaxies at $z\le4$. 
This indicates that towards higher redshift in the EoR, the evolution of $T_{\mathrm{peak}}$ is stronger for the massive galaxies in \flares\ than their low redshift relation. 

In Fig.~\ref{fig: Tdustevo}, we also compare our values to other theoretical and observational results at similar redshifts. 
We compare to observational values from \cite{Strandet2016,Faisst2020b}, with most values from \flares\ in good agreement or otherwise within the constraints. 
There are a few galaxies in \cite{Strandet2016} which show slightly colder $T_{\mathrm{peak}}$ values, in tension with our predictions. 
\flares\ fails in this case to be fairly representative of such cold dusty galaxies. 
It can be seen that our $T_{\mathrm{peak}}$ values are offset from the relation obtained from the \textsc{MassiveFire} simulations in \cite{Liang2019} for $2\le\,z\le6$. The results are in agreement in the region for which the fit was obtained, but would be an overprediction on extrapolation. This difference could be due to the smaller sample size (29 massive galaxies in total) as well as the use of a higher dust-to-metal (DTM=0.4, increasing the optical depth) ratio in that study.
Our result is also very similar to the recent values from the \textsc{Illustris-Tng} \cite[][]{Shen2022} suite of simulations using \skirt, with their reported median $T_{\mathrm{peak}}$ being slightly higher at $z=4,6,8$. 

We can compare our $T_{\mathrm{SED}}$ values to the ones in \cite{Faisst2020b}, since they use the same fitting function as in equation~(\ref{eq: GP-MBB}) (in their work $\alpha$ is fixed at 2.0, since the SED is not constrained blueward of rest-frame $\sim110$ \mumetre). 
The values obtained in that study provide a very reasonable match to our constraints within the median spread, while the other observational results are lower by $\gtrsim10$ K. The measurement from \cite{Jin2019} uses an optically thick MBB that gives dust SED temperatures that are very similar to the ones obtained from equation~(\ref{eq: GP-MBB}) \cite[see Figure~2 in][which shows that an optically-thick MBB gives similar dust temperatures to $T_{\mathrm{SED}}$]{Casey2012}.
One of the values is in very good agreement with our measures for $T_{\mathrm{SED}}$, while the other galaxy at $z\sim5$ in the work has a very cold dust SED temperature. 

For all the other measurements it would be fairer to compare the values to $T_{\mathrm{SED,RJ}}$, as equation~(\ref{eq: MBB-OT}) is used in many of the high-redshift studies to fit for the observed fluxes/luminosities. 
As such our predicted values provide a reasonable match to the observational data from \cite{Strandet2016}, \cite{Hashimoto2019} and \citet[measures the temperature using stacked galaxies with SFR $\ge 10 \, \mathrm{M_{\odot}}$/yr]{Bethermin2020}. 
There are a few exceptions in the observational data that deviate strongly from our predictions. 
For example, \cite{Harikane2020} fit an optically thick MBB to their galaxies at $z\sim5$, and find very cold dust SED temperatures. 
However, in the case of \cite{Bakx2020} the lower limit they provide is very high compared to our predictions, which could be reconciled if using a much high emmisivity index. Some of the large dispersions seen in the dust temperature measures in observations point towards either a wide range of values existing in the diverse populations of galaxies in the early Universe, or an indication of more varied dust grain properties such as their size, shape, and composition that is not captured in the models being employed to study them. 
The \cite{Bouwens2020} fit to the dust SED temperatures (with some of the \citealt{Faisst2020b} peak dust temperature values also being used), which were used in their modified blackbody fitting, does have a few values at $z\ge5$, however the majority of the constraints are from lower redshift (see their fig.~1). The \flares\ dust SED temperatures (referring to $T_{\mathrm{SED,RJ}}$) have slightly higher normalisation compared to their fit, but are in reasonable agreement within the 16-84th percentile spread. 
Similar to our results, the $T_{\mathrm{SED}}$ values in \cite{Shen2022} are also consistently higher than the observationally quoted SED temperatures, with their median being slightly higher than our values. 

It should also be noted that the plotted values and the spread are weighted based on which overdensity region they are from, and thus contribution from the extreme overdensities will be down-weighted. 
Even though they are rare, all the observational values outside the $1\sigma$ scatter are well within the maximum and minimum values of the seen dust temperatures across redshifts in \flares, except for the lower limit measurement from \cite{Bakx2020}. 

%

\section{Conclusions}\label{sec:conc}
We have presented the dust SED properties of galaxies in \flares{}, a suite of zoom simulations that uses the \eagle\ \cite[]{schaye2015_eagle,crain2015_eagle} physics to probe a range of overdensities in the EoR. 
We select massive galaxies ($\gtrsim 10^9$ M$_{\odot}$) in the simulation to make a comprehensive statistical study of galaxies that are accessible to current telescopes. These galaxies are post-processed with the radiative transfer code \skirt\ \cite[]{skirt2015,skirt9} to generate their full SEDs. The dust-to-metal ratios were derived from the fitting function from the dust model implemented in the L-Galaxies SAM \cite[]{Vijayan2019}. We do not calibrate any of the parameters in \skirt\ to produce the SEDs.

Our main findings are as follows:
\begin{enumerate}
	\item The predicted UV LF is in agreement with available observational data. We also compare the IR LF to the observations, finding good agreement at $z=5,6$ for luminosities $<10^{12}$ L$_{\odot}$. We underestimate the number densities of the most IR luminous galaxies at $z=5$. We attribute this mainly to the lack of high star formation rates in the massive galaxies in the simulation, similar to what previous \flares\ or \eagle\ studies have shown \cite[\eg][]{katsianis_evolution_2017,Baes2020,Lovell2021a}. 
	We also underpredict the number of luminous rest-frame 250 \mumetre\ galaxies. However, the observations \cite{Koprowski2017,Gruppioni2020} show discrepancies among each other. 
	The extreme IR objects are biased towards the highest matter overdensities.
	\item The \flares\ IRX-$\beta$ relation for $5\le z\le8$ is consistent with the starburst relation \cite[\eg][]{Meurer1999,Reddy2015} from local redshifts. We see a shift towards the SMC relation \cite[]{Pettini1998} for $z>8$. We see that it is predominantly the lower-mass ($\lesssim10^{9}$ M$_{\odot}$) galaxies that deviate towards this relation. Also galaxies with low-sSFR lie further away from these empirical relations. We see a good match with current available observations, missing a few low-IRX high-$\beta$ galaxies.		
	\item The IRX shows a gradual increase with stellar mass, showing a flattening at high-stellar masses ($\sim10^{10}$ M$_{\odot}$). We do not see any evolution in the normalisation of the median relation with redshift.
	\item We look at the evolution of the peak of the IR emission ($\lambda_{\mathrm{peak}}$) with redshift on properties like the galaxy stellar mass, total IR luminosity and the sSFR. We see flattening of $\lambda_{\mathrm{peak}}$ at high IR luminosity ($\sim10^{11}$ L$_{\odot}$). The $\lambda_{\mathrm{peak}}$ ($T_{\mathrm{peak}}$) - $L_{\mathrm{IR}}$ relation is offset from the observed local relation \cite[]{Casey2018} to lower (higher) values. $\lambda_{\mathrm{peak}}$ strongly correlates with the galaxy sSFR.
	\item Luminosity-weighted dust temperatures (peak dust-temperature: $T_{\mathrm{peak}}$, SED temperature fit from mid-IR powerlaw+MBB: $T_{\mathrm{SED}}$ and SED temperature fit from optically-thin MBB: $T_{\mathrm{SED,RJ}}$) increase with increasing redshift. We find that, for the massive galaxies in \flares, the evolution of $T_{\mathrm{peak}}$ with redshift is stronger than the low-redshift relation obtained from observational \cite[]{Schreiber2018} and other theoretical \cite[]{Liang2019} studies. 
	\item The SED temperatures ($T_{\mathrm{SED}}$ and $T_{\mathrm{SED,RJ}}$) are mostly in agreement with the observational values. However we find a lack of extremely cold temperatures seen in some observations \cite[]{Strandet2016,Jin2019}.
\end{enumerate}

Future observations from many of the planned surveys and observations on \alma{}, \jwst{}, \rst{}, \euclid{}, as well as future IR missions sampling more of the SED will be able to put better constraints on these dust driven properties. In a future work we will explore the dust-continuum sizes of these galaxies.

Through this study as well as previous other works referenced here, there is some evidence in favour of the \eagle\ physics model requiring higher star-formation rates to match some of the observations at high-redshift like the UV LF, IR LF or the sub-mm number counts. However, reconciliation of such limitations must be achieved without losing some of the remarkable successes of the model across the low-redshift Universe. The strive to succeed in this extremely non-trivial challenge has been the goal of all theoretical studies of galaxy formation and evolution. The high-redshift Universe is a regime where \eagle\ as well as other periodic boxes have not been well studied due to its lack of massive galaxies. Studies with \flares\ allows for a statistical exploration of this regime due its novel re-simulation strategy targeting massive overdensities. This will inevitably help to improve theoretical models of galaxy formation and evolution in terms of providing insights into the different feedback mechanisms as well as star formation recipes implemented.

\section*{Acknowledgements}
We thank the anonymous referee for their helpful comments. Thanks to Mark Sargent, Rebecca Bowler, Romeel Dav\`e and Seb Oliver for helpful discussions. We also thank the ALPINE-ALMA collaboration for making their data products public. We also thank Jorge Zavala and Yueying Ni for helpful correspondences. Thanks also to Caitlin Casey for providing their updated IRX data. This work used the DiRAC@Durham facility managed by the Institute for Computational Cosmology on behalf of the STFC DiRAC HPC Facility (www.dirac.ac.uk). The equipment was funded by BEIS capital funding via STFC capital grants ST/K00042X/1, ST/P002293/1, ST/R002371/1 and ST/S002502/1, Durham University and STFC operations grant ST/R000832/1. DiRAC is part of the National e-Infrastructure. We also wish to acknowledge the following open source software packages used in the analysis: \textsc{Numpy} \cite[]{Harris_2020}, \textsc{Scipy} \cite[][]{2020SciPy-NMeth}, \textsc{Astropy} \cite[][]{Astropy2013,Astropy2018}, \textsc{Matplotlib} \cite[][]{Hunter:2007} and WebPlotDigitizer \cite[][]{Rohatgi2020}. 

This manuscript was written during the global Covid-19 pandemic. We would like to acknowledge our privileged positions that provided us the opportunity to pursue research in these tough times, while many members of our global society cannot afford to do that. We would also like to thank everyone behind the development of various Covid-19 vaccines. APV acknowledges the support of his PhD studentship from UK STFC DISCnet. The Cosmic Dawn Center (DAWN) is funded by the Danish National Research Foundation under grant No. 140. CCL acknowledges support from the Royal Society under grant RGF/EA/181016.

\section*{Data Availability}

The data associated with the figures in the paper can be found at \href{https://flaresimulations.github.io/data.html}{https://flaresimulations.github.io/data.html}.



\bibliographystyle{mnras}
\bibliography{ref} 




\appendix

\section{Convergence Tests}\label{sec:conv}
\begin{figure}
	\includegraphics[width=\columnwidth]{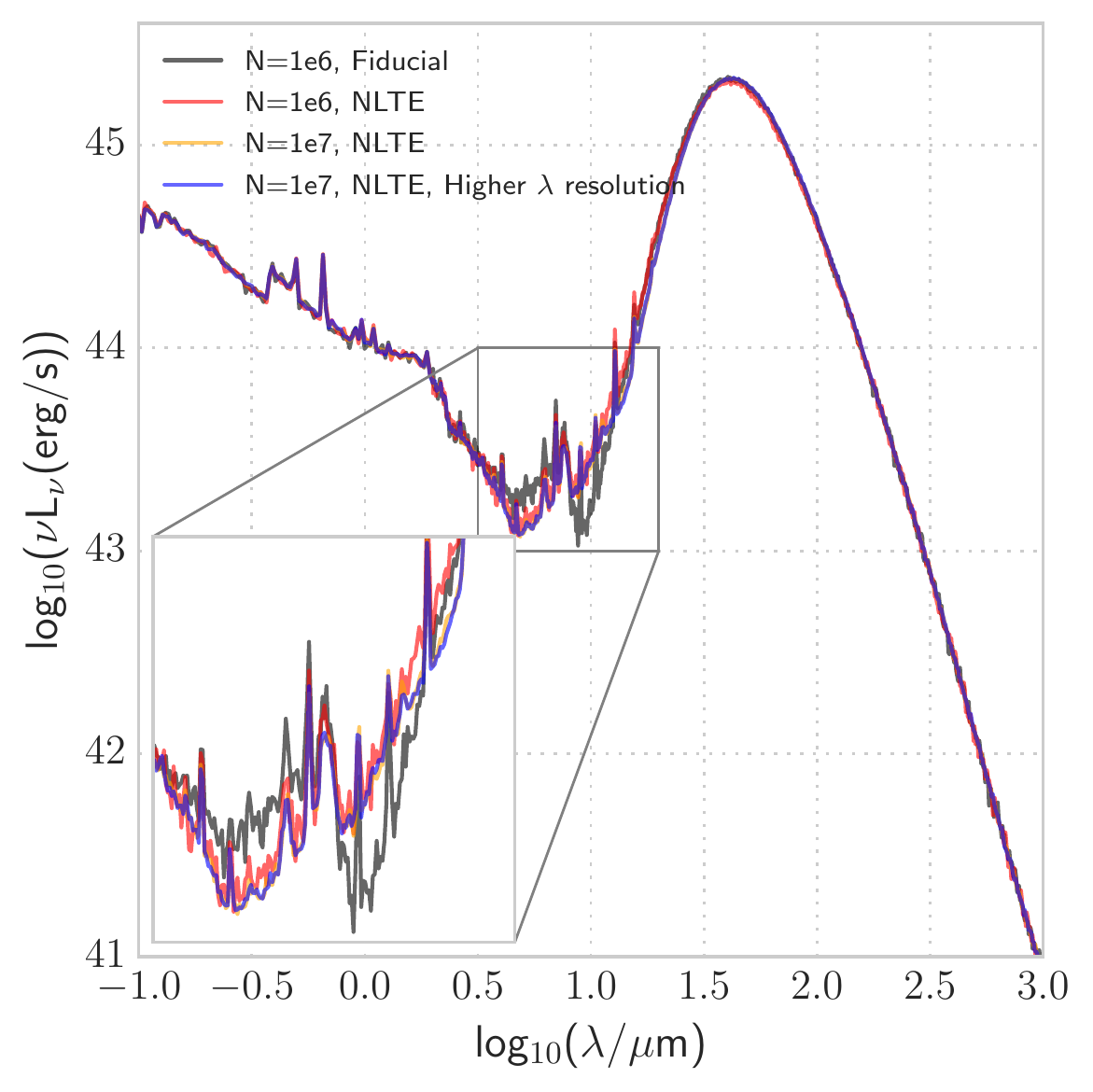}
	\caption{Shows the SED of a galaxy at $z=8$ for a few different configurations of the \skirt\, code. Shown in inset, the mid-IR region. The galaxy has a total IR luminosity of $\sim 10^{11.8}$ L$_{\odot}$ in all the plotted configurations.\label{fig: app_conf1}}
\end{figure}
\begin{figure}
	\includegraphics[width=\columnwidth]{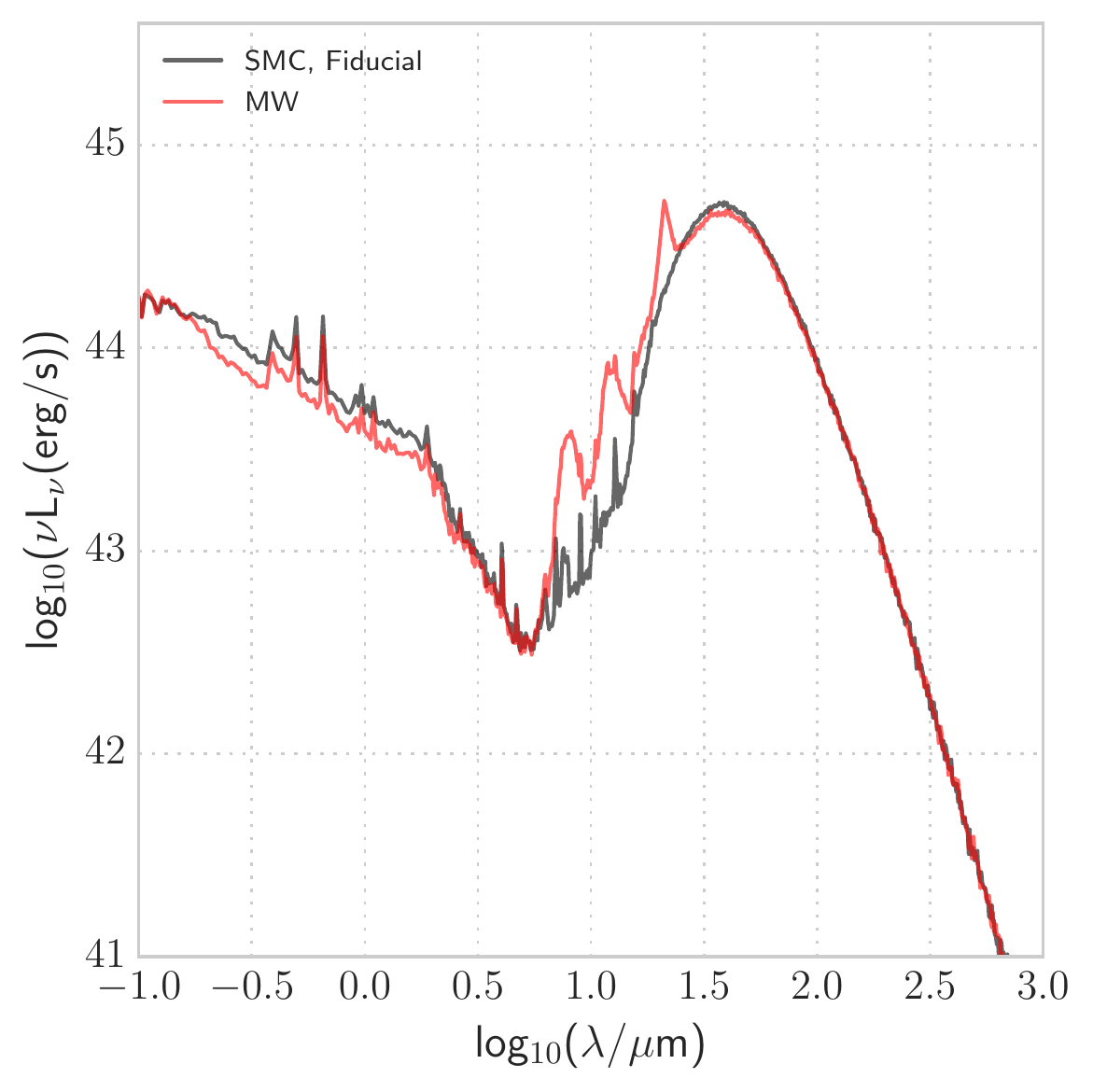}
	\caption{Shows the SED of a galaxy at $z=8$ for the SMC and Milky Way grain distribution. \label{fig: app_conf2}}
\end{figure}
In this section we will test the convergence of our SEDs in relation to some of our \skirt\ parameter choices.
We will first explore how the choice of an increase in the photon number, increasing resolution of our radiation and dust wavelength grid, and adding Stochastic (Non Local Thermal Equilibrium, NLTE) heating to the set-up can change our results. In Fig.~\ref{fig: app_conf1}, we plot the SED of a galaxy at $z=8$ for these different configurations. 
Our higher photon number test runs have 10$^7$ (10 times higher compared to the default run) photons per radiation field wavelength grid, while for the increased wavelength grid set-up we have doubled the number of bins from our default configuration. The total IR luminosity only changes by $\sim0.01$ dex for the galaxy between these choices, while there is even smaller effect in the UV and optical part of the SED. We have checked this for a few other galaxies and find that the changes are similar. There are no noticeable dramatic changes.

We also change our dust grain distribution choice from SMC like to Milky Way (MW) like in Fig.~\ref{fig: app_conf2}. The MW configuration has Poly-Aromatic Hydrocarbons (PAHs) included in the dust grain distribution, which will have an effect on the mid-IR range of the SED. There is a clear indication of an absorption feature near 2175 \AA, due to the bump in the extinction curve for the MW dust distribution. This ultimately leads to $\beta$ values that are always negative. It can be seen from the figure that there is also stronger extinction at wavelengths short of the mid-IR for the MW type, with higher emission in the mid-IR compared to SMC type. 
Next generation instruments that can scan the mid-IR SED at high-redshift are needed to put constraints in this regime, and to aid our understanding of emission from PAHs. 
The change to MW type grain has negligible effect on where the peak of the IR emission is. The total IR luminosity also sees negligible change towards higher values. However, the value of $\beta$ is affected and thus can drive changes in the IRX-$\beta$ plane by making the $\beta$ values more negative.


\section{Effect of AGN on the dust SED}\label{sec:AGN}
\begin{figure*}
	\centering
	\includegraphics[width=\textwidth]{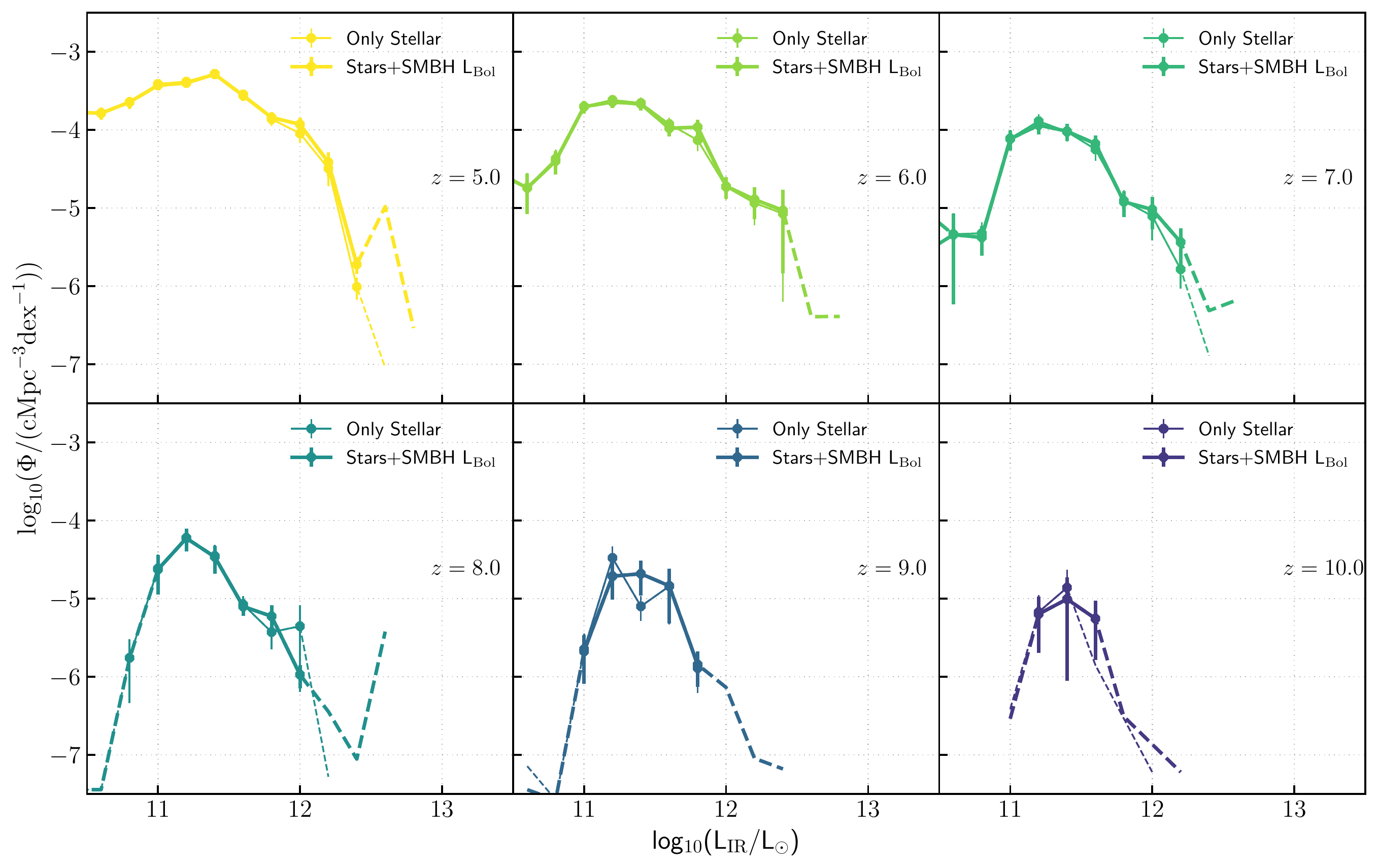}
	\caption{Same as Fig.~\ref{fig: IRLF}, now showing how the total infrared luminosity function changes, if all the energy from the SMBH accretion, as described in equation~(\ref{eq: BH lum}) went into the infrared. This is shown with thicker lines, while the original IR LF is shown by lighter lines.\label{fig: IR+SMBH}}
\end{figure*}
In order to understand the effect AGN have on the observed galaxy SEDs, we perform a simple analysis. For this purpose, we obtain the intrinsic bolometric luminosity of the SMBH in our galaxy sample, and add it to the total IR luminosity of our galaxies. This would represent an upper limit on the total IR luminosity that the galaxy can have from AGN contribution. The SMBH bolometric luminosity is calculated using
\begin{gather}\label{eq: BH lum}
L_{\mathrm{BH, bol}}=\eta\,\frac{\mathrm{d}M_{\bullet}}{\mathrm{d}t}c^2,
\end{gather}	 
where d$M_{\bullet}$/d$t$ is the accretion rate and $\eta$ is the efficiency, assumed to be $0.1$.

In Figure~ \ref{fig: IR+SMBH}, we plot the IR LF with the AGN bolometric luminosity added to the \flares\ galaxies for $z\in[5,10]$. For comparison we also show the IR LF (which only includes stellar reprocessed dust emission) plotted in Figure~\ref{fig: IRLF}
We do not plot any of the observations that were shown in Figure~\ref{fig: IRLF}. 
We can see that there is a small change at the very bright end of the function. However this increase is not enough to reconcile the relation with the observational data. Thus, as explained in \S\ref{sec:res::LF}, the main driver of the difference can be attributed to the lack of more intense star formation activity in massive/bright galaxies in the model or the observations probing a biased region in the Universe as suggested in \cite{Zavala2021}.

\section{Excluding birth cloud emission}\label{sec:BC emission}

\begin{figure*}
	\centering
	\includegraphics[width=0.7\textwidth]{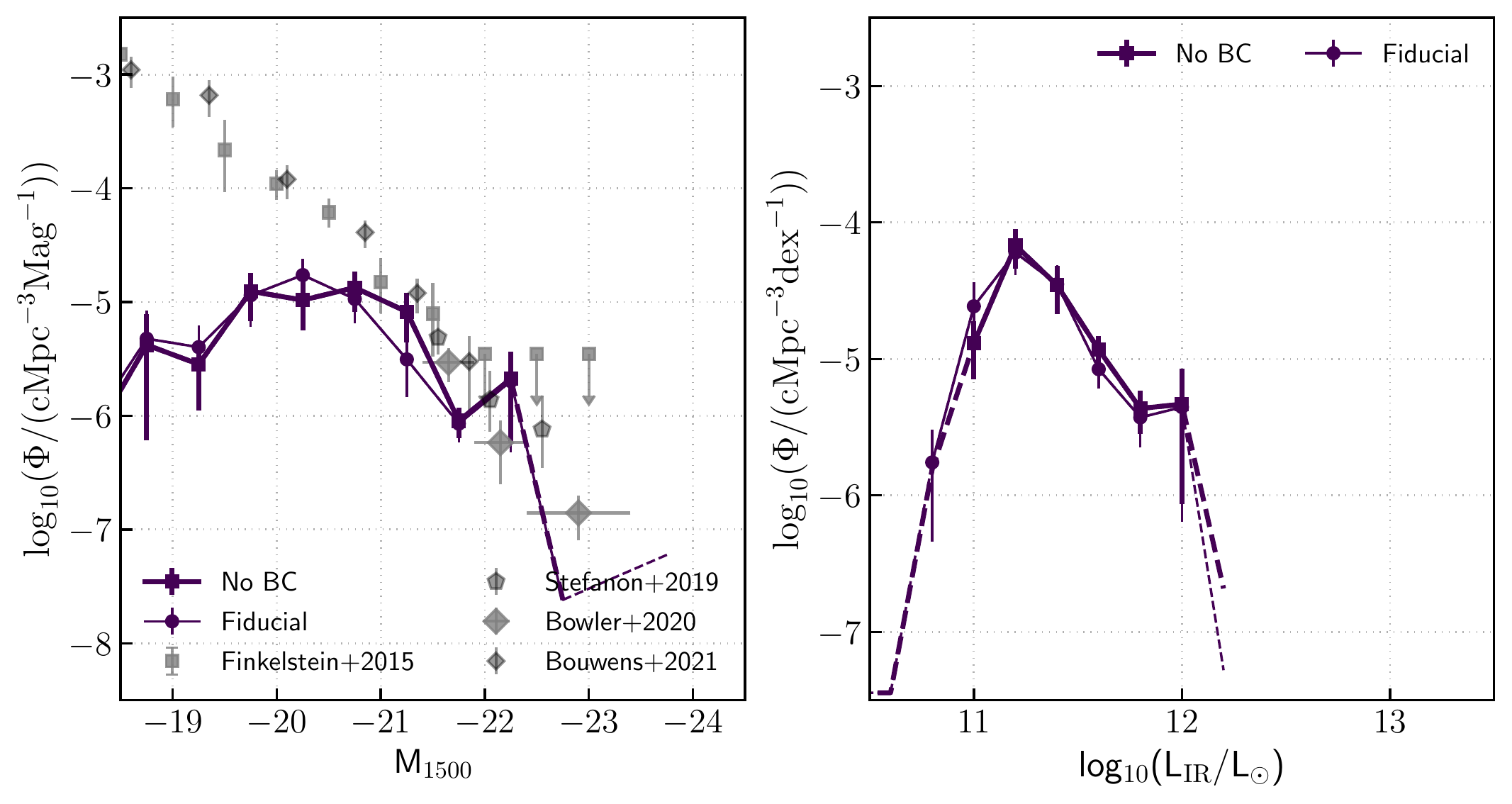}
	\caption{Plot compares the UV (left panel) and IR (right panel) LF with (thin line) and without (thick line) the birth cloud attenuation for $z=8$. 
	Observational data, same as in Figure~\ref{fig: UVLF}, is also plotted in the left panel. \label{fig: noBC compare}}
\end{figure*}

To see the effect of birth cloud emission in the model, we ran \skirt\ for galaxies only at $z=8$ (in order to drastically reduce the computational costs, since there are only fewer galaxies at $z=8$ compared to lower redshifts), by treating the young star-forming particles as regular star particles, without the added dust extinction implemented in the MAPPINGS III template due to the birth cloud. We now model both these radiation sources using BPASS, ignoring this extra extinction. 
We plot the UV and IR LF of the results in Figure~\ref{fig: noBC compare}, as well as compare to our fiducial set-up. It can be seen that the UV LF is within the scatter at each bin. Similarly, in case of the IR LF at $z=8$, there is only very negligible change. The small impact on both these functions is due to the low metallicity of these systems, owing to them being at extremely high-redshift. It is expected that with the increase in metallicity of systems at lower redshifts, birth cloud attenuation will have more of an influence. 

\section{T$_{\mathrm{\lowercase{peak}}}$ and galaxy main-sequence}\label{sec: tpeak and MS}
\begin{figure*}
	\centering
	\includegraphics[width=\textwidth]{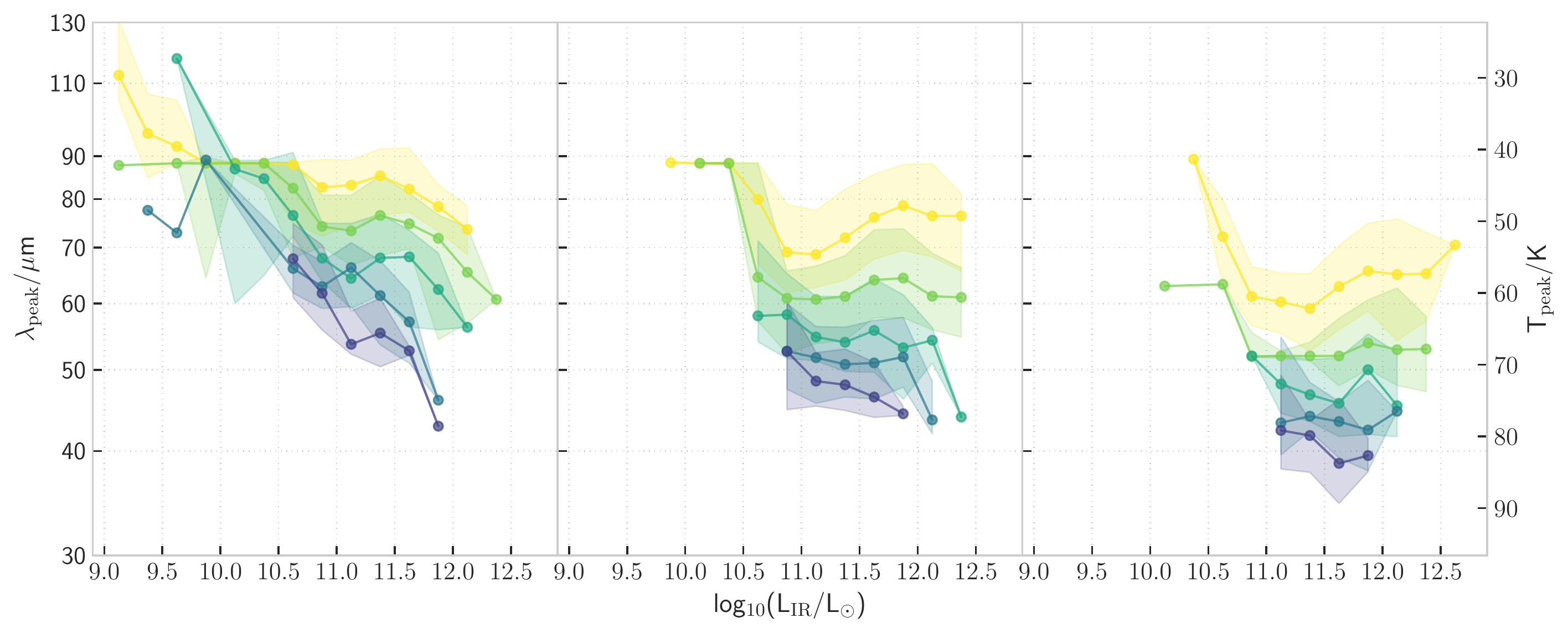}
	\caption{We show the variation of $\lambda_{\mathrm{peak}}$ (corresponding $T_{\mathrm{peak}}$ values is shown on the right y-axis) with the IR luminosity for $z\in[5,9]$. The panels represent galaxies with suppresed star-formation (left), galaxies on the main-sequence (middle) and starburst galaxies (right). \label{fig: lampeakms}}
\end{figure*}

To better understand the relation between $\lambda_{\mathrm{peak}}$ and the sSFR, we separate the \flares\ galaxies into 3 groups, based on an evolving piecewise fit to the stellar mass-SFR relation presented in \citet[][see \S3.4 in the work, equations 11 and 12]{Lovell2021a}. The groups have been classified based on their deviation from the piecewise fit.
The 3 groups have been plotted in Figure~\ref{fig: lampeakms}, with the median $\lambda_{\mathrm{peak}}$ as a function of the IR luminosity. These groups are galaxies
\begin{itemize}
	\item below $1\sigma$ (left panel in Figure~\ref{fig: lampeakms}) from the fit, which can be classified to include the green valley and the passive galaxies,
	\item within $1\sigma$ (middle panel in Figure~\ref{fig: lampeakms}) of the fit relation on either sides, termed the main-sequence, and
	\item above $1\sigma$ (right panel in Figure~\ref{fig: lampeakms}) of the fit, which can be termed as starbursts. 
\end{itemize}
We only show in Figure~\ref{fig: lampeakms} the relation for $z\in[5,9]$, since the star-forming sequence fit could not be constrained properly at $z=10$ \cite[see][]{Lovell2021a}.

It can be seen from Figure~\ref{fig: lampeakms} that the shape of the relation between $\lambda_{\mathrm{peak}}$ and $L_{\mathrm{IR}}$ is different for the 3 groups. 
Galaxies in the green valley/passive regime (left panel) show a consistent decrease (increase) in $\lambda_{\mathrm{peak}}$ ($T_{\mathrm{peak}}$) with $L_{\mathrm{IR}}$. 
This is mainly due to the smaller dust content within these galaxies and thus there is a direct correlation between the increase in dust-temperature and $L_{\mathrm{IR}}$. The other two groups exhibit a flat relationship with $L_{\mathrm{IR}}$, due to their high dust content and thus the hot dust being optically thick, similar to that seen in some observations at high-redshift \cite[\eg][]{Cortzen2020}. 
The starburst galaxies have a lower median $\lambda_{\mathrm{peak}}$ due to their higher sSFRs. At $z=5$, towards high $L_{\mathrm{IR}}$ there is also a hint of increasing $\lambda_{\mathrm{peak}}$ values, indicating the rapid build up of dust in these extreme objects. 

\section{Comparison with line-of-sight model}\label{sec:los_comp}
\begin{figure*}
	\centering
	\includegraphics[width=\textwidth]{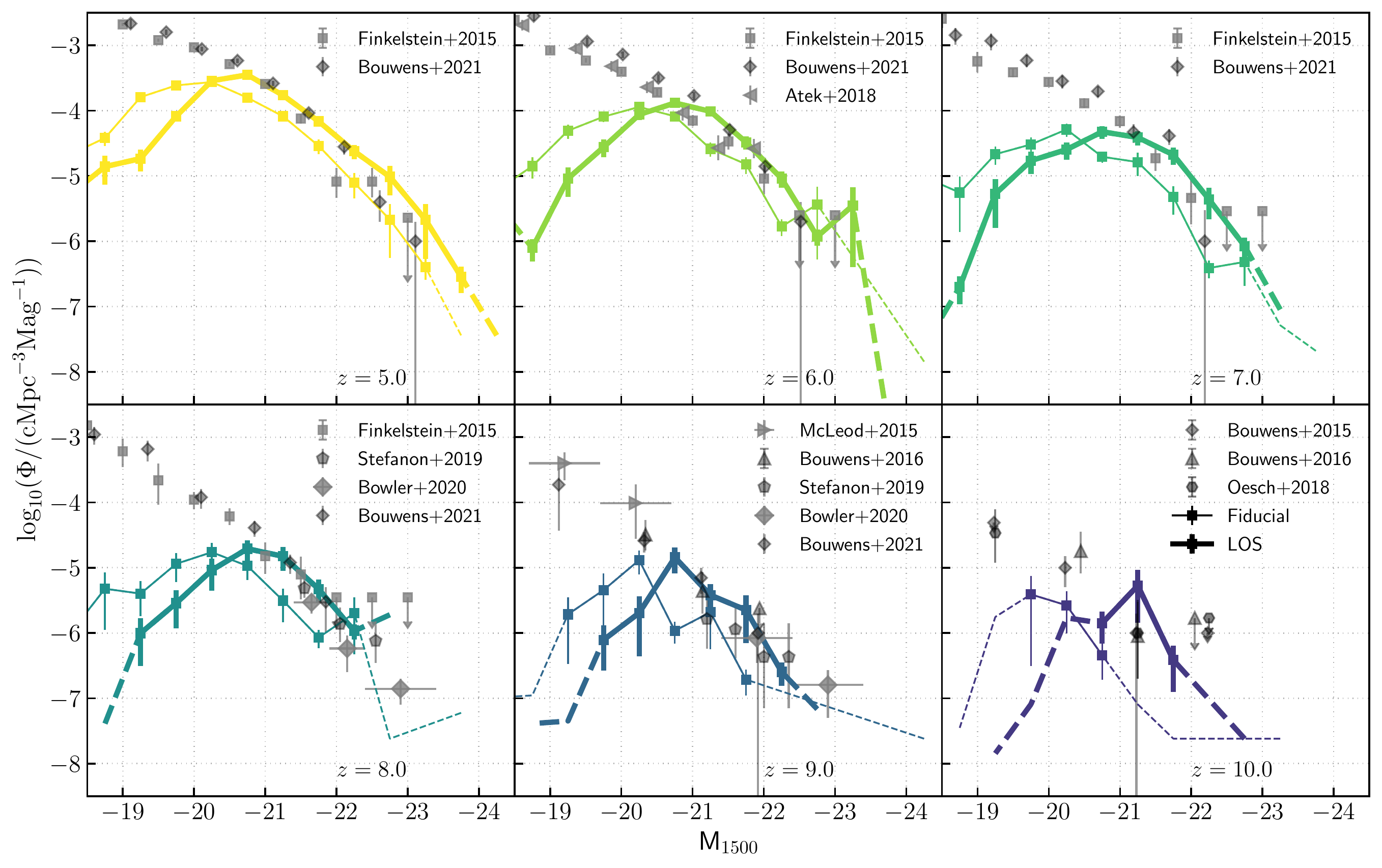}
	\caption{Same as Fig.~\ref{fig: UVLF}, but here we also include the UV LF obtained from the LOS dust extinction model (thicker lines) implemented in \protect\cite{Vijayan2021} on the \flares\, galaxies. \label{fig: LOS compare}}
\end{figure*}

In this section we will compare the UV luminosity obtained from our \skirt\ modelling here to the line-of-sight (LOS) dust model we implemented on the same galaxies in \cite{Vijayan2021}. 
In that work we assumed a dust attenuation curve and modelled the dust attenuation parameters for the old stars and the young stars to find a good match to the $z=5$ UV LF and the UV-$\beta$ relation, as well as observations of the [O\textsc{iii}]+H$\beta$ EW relations at $z=8$. 
This method is computationally much faster, and allows for more flexibility in the modelling, allowing you to explore changes and their effects more easily. 
However, it can not treat certain phenomena, such as the scattering of light away from the LOS or dust self-absorption, as these processes are dependent on the chosen extinction curve. 

We compare the UV luminosity of the galaxies selected in this work, using the two dust models in Fig.~\ref{fig: LOS compare}, for $z\in[5,10]$. We can see that the values obtained in this work are systematically lower than the ones from the LOS model by$\sim0.4$dex. This is mainly due to the lower dust optical depth (parameterised by the $\kappa$ parameters) along the LOS adopted in that study compared to this work. This by construction matches the observations that the model was calibrated for. We have already explained one of the reasons for the slightly lower number densities in \S\ref{sec:res::LF} for the non-calibrated model presented here.

\bsp	
\label{lastpage}
\end{document}